\documentclass[twocolumn,aps,prd,   
               preprintnumbers,numbers,sort&compress,
               nofootinbib,
               showpacs,
               floatfix,
               colorlinks,
               linkcolor=blue,
               citecolor=blue,
               10pt]{revtex4-1}

\usepackage{graphicx,amsmath,amssymb,bm}
\usepackage{psfrag}
\usepackage{feynmp}
\usepackage{hyperref}
\usepackage{enumitem}

\newcommand{\exclude}[1]{}

\newcommand{\beq}{\begin{equation}}
\newcommand{\eeq}{\end{equation}}
\newcommand{\be}{\begin{eqnarray}}
\newcommand{\ee}{\end{eqnarray}}

\def\+{\dagger}
\def\la{\langle}
\def\ra{\rangle}
\def\<{\langle}
\def\>{\rangle}

\begin{document}

\title{Oblique Confinement at $\theta\neq 0$ in  weakly coupled gauge theories with deformations.}

\author{ Mohamed M. Anber}
\email{manber@lclark.edu}
\affiliation{Department of Physics, Lewis \& Clark College, 
Portland, OR 97219, USA}
\author{ Ariel R. Zhitnitsky} 
\email{arz@phas.ubc.ca}
\affiliation{Department of Physics \& Astronomy, University of British Columbia, Vancouver, B.C. V6T 1Z1, Canada}


\begin{abstract}
 
The main focus of this work is to test the ideas related to the oblique confinement in a theoretically controllable manner using the ``deformed QCD'' as a toy model. We explicitly show  that the oblique confinement in the weakly coupled gauge theories  emerges  as a result of condensation   of $N$ types of monopoles  shifted by the phase $\exp(i\frac{\theta+2\pi m}{N})$ in Bloch type construction. It should be contrasted with conventional and commonly accepted viewpoint that the confinement at $\theta\neq 0$ is due to the  condensation of the electrically charged dyons which indeed normally  emerge in the systems with $\theta\neq 0$ as a result of  Witten's effect.  We explain  the basic reason  why the  ``dyon" mechanism  does not materialize-- it is because the Witten's effect holds  for a static magnetic monopole treated as an external source. It should be contrasted with  our  case when   $N$- types of  monopoles are not static, but rather the  dynamical degrees of freedom which fluctuate and themselves determine the ground state of the system.

 \end{abstract}

\maketitle

\section{Introduction and motivation} \label{introduction}

A study of the the QCD vacuum state in the strong coupling regime is the prerogative of numerical Monte Carlo lattice computations.
However, a number of very deep and fundamental questions about the QCD vacuum structure can be addressed and, more importantly, answered using some simplified versions of QCD. 
In the present paper, we study a set of questions intimately connected to the ground state with $\theta\neq 0$.  We use the so-called ``deformed QCD'' and similar toy models wherein we can work analytically\footnote{\label{terminology}We use one and the same term ``deformed QCD" model 
for systems with and without quarks.  We hope it  does not confuse the readers as  specific   description of the system should be obvious from the context of the discussions.}.
These models   belong to the class  of the  weakly coupled gauge theory, which however preserves many essential elements expected for true QCD, such as confinement, degenerate topological sectors, proper $\theta$ dependence, etc.
This allows us to study difficult and nontrivial features, particularly related to vacuum structure at $\theta\neq 0$, in an analytically tractable manner.

The $\theta$ dependence in the system is intimately related to the presence of the metastable states which always accompany the gauge systems
even at $\theta=0$. 
The fact that some high energy metastable vacuum states must be present in a gauge theory system in the large $N$ limit has been known for quite some time \cite{Witten:1980sp}. A similar conclusion  also  follows from the  holographic description of QCD as originally discussed in \cite{Witten:1998uka}.  Therefore, 
the understanding of the microscopical description  of the ground state at $\theta\neq 0$ in terms  of the monopoles (in the ``deformed QCD'' and other toy models as will be discussed in the present work) inevitably requires the   microscopical understanding of these metastable states   as both constructions, the $\theta\neq 0$ states and the metastable states at $\theta=0$,  must be described simultaneously in terms of the same degrees of freedom and in terms of the same fundamental gauge configurations.   

 \subsection{$\theta\neq 0$: Phenomenological motivation}\label{phenomenology} 
 
The questions  being  addressed in the present work, as highlighted above, are very deep and fundamental problems of the strongly coupled gauge theory. 
One could naively think that these problems with $\theta\neq 0$ are  pure academic questions which have no physics applications, observable consequences or any phenomenological significance    as it is known that  $\theta=0$ with extremely  high accuracy  in our  Universe at present time.  However we want to emphasize  here that, in fact,  the problems highlighted above were  largely motivated by an attempt to understand  the QCD transition in the early Universe when $\theta$ was not identically zero, but rather was slowly  relaxing  to zero field as a result of the axion dynamics, see original papers \cite{Peccei:1977ur,Weinberg:1977ma,Wilczek:1977pj,Shifman:1979if,Kim:1979if,Dine:1981rt,Zhitnitsky:1980tq} and   review articles \cite{vanBibber:2006rb,Asztalos:2006kz,Raffelt:2006cw,Sikivie:2009fv,Rosenberg:2015kxa,Graham:2015ouw,Ringwald:2016yge} on the theory of axion and recent advances in the axion search experiments. 

The recent lattice studies \cite{Kitano:2015fla,Bonati:2015vqz,Borsanyi:2016ksw,Petreczky:2016vrs} addressing related questions 
on  the axion dynamics during the QCD transition  essentially  are capable to compute the correlation functions, such as the topological susceptibility (\ref{chi})  at $\theta=0$,
while the gauge configurations at $\theta\neq 0$ are not accessible by conventional lattice methods. 
 The study of the dynamics of the system at $\theta\neq 0$ represents a very challenging technical problem as a result of the so-called ``sign problem".   Therefore, at present time the lattice studies can provide limited information on microscopical dynamics  of the strongly coupled gauge theories at finite 
 $\theta\neq 0$ \cite{DElia:2012pvq,DElia:2013uaf}, specially the regions in vicinity of $\theta\simeq \pi$ when the level-crossing phenomena is expected to occur, and  metastable states become almost degenerate with the ground vacuum states.

At the same time, a precise understanding of the structure of the ground state at $\theta\neq 0$ and its microscopical description during this complicated time evolution  plays a crucial role in computations of the axion production rate, possible formation of the axion domain  walls\footnote{\label{DW}In particular, the so-called $N=1$ domain walls corresponding to the interpolation of the axion $\theta(x)$ field between topologically distinct but physically identical states $\theta=0$ and $\theta=2\pi$  will  inevitably form   due to the $2\pi$ periodicity in $\theta$ and presence of the metastable states mentioned above. The formation of such kind of $N=1$ domain walls happen   irrespectively whether the inflation occurs before or after the PQ phase transition, see comments in \cite{Liang:2016tqc,Ge:2017ttc}.}, possible role of the metastable states (which inevitably are present in the system as will be argued in this work), and many other related questions which essentially determine the dark sector of the Universe at present time.

 The main claim  which will be advocated in  the present work is that the microscopical description of the oblique confinement at $\theta\neq 0$  
is due to the  condensation of the same fractionally charged monopoles in ``deformed QCD" model which are responsible for the confinement at $\theta=0$.
The same microscopic description remains  also valid for the metastable vacuum states which are always present in gauge theories. The only modification which occurs in the description  for metastable states and $\theta\neq 0$ states   is that the vacuum expectation value of the magnetization operator gets 
shifted by the phase 
$\exp(i\frac{\theta+2\pi m}{N})$ in Bloch type construction.
 We  reiterate the same claim as follows: we do not see any     room within our framework for the commonly accepted   ``dyon mechanism" for the oblique confinement, speculated long ago by t`Hooft \cite{tHooft:1981bkw} when the electrically charged dyons condense.    
  
  We conjecture that this picture we have just described  holds in strongly coupled regime as well, not only in the weakly coupled ``deformed QCD" model.   
We present few arguments supporting this conjecture in the next subsection.

 \subsection{Smooth transition between weakly coupled and strongly coupled regimes.}\label{smooth}
When some deep questions are studied in a simplified version of a theory, there is always a risk that some effects which emerge in the simplified version of the theory could be just artifacts of the approximation, rather than genuine consequences of the original underlying theory.
Our present studies in this work  using the ``deformed QCD'' and other toy models are  not free from this difficulty of possible  misinterpretation of artifacts as inherent features of underlying QCD.
Nevertheless, there are a few strong arguments suggesting that we indeed study some intrinsic features of the system rather than some artificial effects.
The first argument has been presented in the original paper on ``deformed QCD'' \cite{Unsal:2008ch} where it has been argued that this model describes a smooth interpolation between strongly coupled QCD and the weakly coupled ``deformed QCD'' without any phase transition.
In addition, there are a few more arguments based on  previous experience \cite{Unsal:2008ch,Poppitz:2011wy,Thomas:2011ee,Anber:2011gn,Unsal:2012zj,Poppitz:2012sw,Thomas:2012tu,Poppitz:2012nz,Zhitnitsky:2013hs,Anber:2013sga,Anber:2013doa,Anber:2014lba,Anber:2017pak,Bhoonah:2014gpa,Aitken:2017ayq} with the ``deformed QCD'' and other toy models which also strongly suggest that we indeed study some intrinsic features of QCD rather than some artifacts of the deformations.

Most of the arguments, with very few exceptions,  from  the previous  studies  \cite{Unsal:2008ch,Poppitz:2011wy,Thomas:2011ee,Anber:2011gn,Unsal:2012zj,Poppitz:2012sw,Thomas:2012tu,Poppitz:2012nz,Zhitnitsky:2013hs,Anber:2013sga,Anber:2013doa,Anber:2014lba,Anber:2017pak,Bhoonah:2014gpa}  of the system 
which are  related to  the $\theta$ dependent physics     are purely analytical  in nature as
they cannot be    independently verified or  tested by using some other means, such as the numerical  lattice simulations. Fortunately, some of the observables, such as the topological susceptibility   $\chi$ defined as 
\be
\label{chi}
\chi =\frac{\partial^2 E_{\rm vac}(\theta)}{\partial \theta^2}|_{\theta=0}=\lim_{k\rightarrow 0}\int e^{ikx}d^4 x \< q(x) , q(0) \>
\ee 
with  $q(x)$ being  the topological density operator, are highly sensitive to the $\theta$ behaviour  even at $\theta=0$ because $\chi$  measures the response of the system with respect to the insertion of the external parameter $\theta$ as one can see from the definition (\ref{chi}). 
 What is more important is that the topological susceptibility $\chi$   can be    also  studied on the lattice at $\theta=0$.
  
The topological susceptibility   $\chi$ has been introduced into the theory long ago \cite{Witten:1979vv,Veneziano:1979ec,DiVecchia:1980yfw} in a course of studies related to the resolution of the $U(1)_A$ problem in QCD in the large $N$ limit. As a result of its fundamental importance for the phenomenological particle physics  the topological susceptibility   $\chi$   has been extensively studied in lattice numerical simulations.  The computations \cite{Thomas:2011ee,Zhitnitsky:2013hs}  of the topological susceptibility in the ``deformed QCD''  model is perfectly consistent with the lattice results, including some extremely nontrivial features  related to the ``wrong sign" of 
the contact term\footnote{It is known  that  the  contact term with a positive sign (in the Euclidean formulation) in  $\chi$ is required for the resolution of the $U(1)_A$ problem \cite{Witten:1979vv,Veneziano:1979ec,DiVecchia:1980yfw}.
At the same time, any physical propagating degrees of freedom must contribute with a negative sign, see \cite{Thomas:2011ee} with details.
In \cite{Witten:1979vv} this positive contact term has been simply postulated while in \cite{Veneziano:1979ec,DiVecchia:1980yfw} an unphysical Veneziano ghost was introduced into the system to saturate this term with the ``wrong'' sign in the topological susceptibility. This entire, very non-trivial framework, has been successfully confirmed by a number of independent lattice computations and precisely reproduced in ``deformed QCD" model.    In addition, one can explicitly see how the Veneziano ghost postulated in \cite{Veneziano:1979ec,DiVecchia:1980yfw} is explicitly expressed in terms of auxiliary topological fields which saturate the contact term   in this model \cite{Zhitnitsky:2013hs}. One can also see that the  $\eta'$ becomes massive in this theory as a result of mixture of a ``would be" Goldstone field with auxiliary topological fields which saturate the contact term in (\ref{chi}).}  and exact cancellation (in the chiral limit) of the contact term with the ``wrong sign" with  physical term  in agreement with the Ward Identities, as described in the original papers  \cite{Thomas:2011ee,Zhitnitsky:2013hs}.

Fortunately enough, there is still one more analytical study in the small circle limit that sheds light on the nature of the phase structure of gauge theories at $\theta \neq 0$. In \cite{Anber:2013sga} a conjectured continuity between mass deformed ${\cal N}=1$ super Yang-Mills on a small circle and pure Yang-Mills at finite temperature was exploited to study the behavior of the thermal phase transition in the latter theories as a function of $\theta$. According to this conjecture, quantum phase transitions in mass deformed ${\cal N}=1$ on $\mathbb R^3\times \mathbb S^1$  are analytically connected to thermal phase transitions in pure Yang-Mills \cite{Poppitz:2012sw,Anber:2014lba}. Thus, one can perform all computations in the small circle limit, where the theory is under analytical control, and then extract conclusions about the strongly coupled theories. It was found in \cite{Anber:2013sga} that the deconfining temperature of any $SU(N)$ gauge theory decreases as $\theta$ increases and also the strength of the first order transition increases with $\theta$. This is in accordance with the lattice simulations that were performed for small $\theta$ in strongly coupled theories \cite{DElia:2012pvq,DElia:2013uaf} and arrived at the same conclusions of \cite{Anber:2013sga}.

We conclude this subsection  with the following generic comment. 
All the features related to the $\theta$ dependence which are known to be present in strongly coupled regime also emerge in the weakly coupled ``deformed QCD''  and other toy models.  
Therefore, we interpret such nice agreement  as a strong argument supporting our conjecture  that these  models properly describe, at least qualitatively, the microscopical features related to the   $\theta$ dependent effects  in the strongly coupled gauge theories.

\subsection{The relation to ${\cal N}=2$   Seiberg Witten model and the structure of  the paper}

Our presentation is organized as follows.
We start in section \ref{deformedqcd} by reviewing a simplified (``deformed'') version of QCD which, on one hand, is a weakly coupled gauge theory wherein computations can be performed in theoretically controllable manner.
On other hand, this deformation preserves all the elements relevant to our study such as confinement, degeneracy of topological sectors, nontrivial $\theta$ dependence,  and other crucial aspects pertinent to the study of the oblique confinement for metastable states and $\theta\neq 0$ states.
In section \ref{classification} we explain the classification of the $\theta$ states while 
in  section \ref{oblique}
we explicitly show that oblique confinement in this model is due to the {\it identically 
same} fractionally charged monopoles which are responsible for the confinement at $\theta=0$. 

This is obviously an expected result especially in view of the arguments presented above   suggesting that this result holds in strongly coupled regime as well due to the  smooth transition between the weakly coupled  ``deformed QCD" and strongly coupled QCD realized in nature.  At the same time the common lore in the community is that the oblique confinement at $\theta\neq 0$ is a result of condensation of the electrically charged dyons which emerge as a result of the Witten's effect \cite{Witten:1979ey}. This common lore is mostly based on analysis of the ${\cal N}=2$   Seiberg Witten model where the dyons are known to be part of   spectrum.
Therefore, it is indeed a quite natural   assumption  that these dyons will condense at $\theta\neq 0$, similar to the monopole's condensation  in the original Seiberg Witten model at $\theta=0$.

Motivated by these  arguments we turn to   ${\cal N}=2$   Seiberg Witten model with the goal to understand 
   the nature of the oblique confinement at $\theta\neq 0$ in  SUSY gauge theories and its relation to studies in ``deformed QCD" model presented in section  \ref{oblique}. We start, in Section  \ref{dyons-static} by  reviewing the  ${\cal N}=2$  SUSY model  defined on $\mathbb R^4$ with emphasize  on the structure of the conventional static dyons and the monopoles in this model. As our goal is to understand the role of these particles in the confinement mechanism at $\theta\neq 0$ and the relation with oblique confinement in ``deformed QCD" model, we formulate ${\cal N}=2$  SUSY model    on $\mathbb R^3\times  \mathbb S^1$ 
 in section \ref{dyons-dynamics} and show that the nonperturbative spectrum of the theory on a sufficiently small circle consists of a tower of monopoles with higher winding numbers.  In section \ref{witten} we explain Witten's effect \cite{Witten:1979ey} in the context of this work, i.e. we explain that the static magnetic monopoles, i.e. 't Hooft lines, indeed become the dyons in the presence of $\theta\neq 0$. However, the magnetic monopoles which play the key role in the confinement mechanism are not static, but rather the  dynamical degrees of freedom which fluctuate and themselves determine the ground state of the system. In the former case the monopoles become the dyons, while in the later  case  they remain pure monopoles with zero electric charges. 

Throughout this work we use the word {\it dyon-particles}  to mean {\it genuine particles (solitons) that carry both electric and magnetic charges}. They are genuine in the sense that they sweep time-like worldlines. We also use {\it dyon-instantons} to mean {\it pseudo-particles that carry both electric and magnetic charges}. They are pseudo since they are only {\it instantaneous events in the Euclidean space} and do not sweep worldlines.  Dyons with zero electric charges are monopoles; these are either monopole-particles or monopoles-instantons.  The words {\it dyons} or {\it monopoles} will be used to mean either particles or instantons when the distinction is either not important or understood from the context.
 
\section{``Deformed QCD" model} \label{deformedqcd}

Here we overview the ``center-stabilized'' deformed Yang-Mills developed in \cite {Unsal:2008ch}. In this section and in Sections \ref{classification} and \ref{classification} we use the words monopoles and dyons to mean monopole-instantons and dyon-instantons, respectively. 
In the deformed theory an extra ``deformation'' term is put into the Lagrangian in order to prevent the center symmetry breaking that characterizes the QCD phase transition between ``confined'' hadronic matter and ``deconfined'' quark-gluon plasma, thereby explicitly preventing that transition.
Basically the extra term describes a potential for the order parameter. 
The basics of this model are reviewed in this section, while in section \ref{classification} we classify the metastable states which are  inherent elements of the system.

We start with pure Yang-Mills (gluodynamics) with gauge group $SU(N)$ on the manifold $\mathbb{R}^{3} \times S^{1}$ with the standard action
\be \label{standardYM}
	S^{YM} = \int_{\mathbb{R}^{3} \times S^{1}} d^{4}x\; \frac{1}{2 g^2} \mathrm{tr} \left[ F_{\mu\nu}^{2} (x) \right],
\ee
and add to it a deformation action,
\be \label{deformation}
	\Delta S \equiv \int_{\mathbb{R}^{3}}d^{3}x \; \frac{1}{L^{3}} P \left[ \Omega(\mathbf{x}) \right],
\ee 
built out of the Wilson loop (Polyakov loop) wrapping the compact dimension
\be \label{loop}
	\Omega(\mathbf{x}) \equiv \mathcal{P} \left[ e^{i \oint dx_{4} \; A_{4} (\mathbf{x},x_{4})} \right].
\ee
The parameter  $L$ here is the length of the compactified dimension which is assumed to be small. 
The coefficients of the polynomial $P \left[ \Omega(\mathbf{x}) \right]$ can be suitably chosen such that the deformation potential (\ref{deformation}) forces unbroken symmetry at any compactification scales.
At small compactification $L$ the gauge coupling is small so that the semiclassical computations are under complete theoretical control \cite{Unsal:2008ch}.

As described in \cite{Unsal:2008ch}, the proper infrared description of the theory is a dilute gas of $N$ types of monopoles, characterized by their magnetic charges, which are proportional to the simple roots and affine root $\alpha_{a} \in \Delta_{\mathrm{aff}}$ of the Lie algebra of the gauge group $U(1)^{N}$.
For a fundamental monopole with magnetic charge $\alpha_{a} \in \Delta_{\mathrm{aff}}$ (the affine root system), the topological charge is given by
\be \label{topologicalcharge}
	Q = \int_{\mathbb{R}^{3} \times S^{1}} d^{4}x \; \frac{1}{16 \pi^{2}} \mathrm{tr} \left[ F_{\mu\nu} \tilde{F}^{\mu\nu} \right]
		= \pm\frac{1}{N},
\ee
and the Yang-Mills action is given by
\be \label{YMaction}
	S_{YM} = \int_{\mathbb{R}^{3} \times S^{1}} d^{4}x \; \frac{1}{2 g^{2}} \mathrm{tr} \left[ F_{\mu\nu}^{2} \right] = \frac{8 \pi^{2}}{g^{2}} \left| Q \right|.
\ee
The $\theta$-parameter in the Yang-Mills action can be included in the conventional way,
\be \label{thetaincluded}
	S_{\mathrm{YM}} \rightarrow S_{\mathrm{YM}} + i \theta \int_{\mathbb{R}^{3} \times S^{1}} d^{4}x\frac{1}{16 \pi^{2}} \mathrm{tr}
		\left[ F_{\mu\nu} \tilde{F}^{\mu\nu} \right],
\ee
with $\tilde{F}^{\mu\nu} \equiv \epsilon^{\mu\nu\rho\sigma} F_{\rho\sigma}/2$.

The system of interacting monopoles, including the $\theta$ parameter, can be represented in the dual sine-Gordon form as follows \cite{Unsal:2008ch}
\be
\label{thetaaction}
	S_{\mathrm{dual}} &=& \int_{\mathbb{R}^{3}}  d^{3}x \frac{1}{2 L} \left( \frac{g}{2 \pi} \right)^{2}
		\left( \nabla \bm{\sigma} \right)^{2} \nonumber \\
	    &-& \zeta  \int_{\mathbb{R}^{3}}  d^{3}x \sum_{a = 1}^{N} \cos \left( \alpha_{a} \cdot \bm{\sigma}
		+ \frac{\theta}{N} \right), 
\ee
where $\zeta$ is magnetic monopole fugacity which can be explicitly computed in this model using the conventional semiclassical approximation.
The $\theta$ parameter enters the effective Lagrangian (\ref{thetaaction}) as $\theta/N$ which is the direct consequence of the fractional topological charges of the monopoles (\ref{topologicalcharge}).
Nevertheless, the theory is still $2\pi$ periodic.
This $2\pi$ periodicity of the theory is restored not due to the $2\pi$ periodicity of Lagrangian (\ref{thetaaction}) as  it was (incorrectly) claimed  in the original  reference \cite{Unsal:2008ch}.  
Rather, it is restored as a result of summation over all branches of the theory when the levels cross at $\theta=\pi (mod ~2\pi)$ and one branch replaces another and becomes the lowest energy state as presented in \cite{Thomas:2011ee}.
   
The dimensional parameter which governs the dynamics of the problem is the Debye correlation length of the monopole's gas, 
\be \label{sigmamass}
	m_{\sigma}^{2} \equiv L \zeta \left( \frac{4\pi}{g} \right)^{2}.
\ee
The average number of monopoles in a ``Debye volume'' is given by
\begin{equation} \label{debye}
	{\cal{N}}\equiv	m_{\sigma}^{-3} \zeta = \left( \frac{g}{4\pi} \right)^{3} \frac{1}{\sqrt{L^3 \zeta}} \gg 1,
\end{equation} 
The last inequality holds since the monopole fugacity is exponentially suppressed, $\zeta \sim e^{-1/g^2}$, and in fact we can view (\ref{debye}) as a constraint on the region of validity where semiclassical approximation is justified. This parameter ${\cal N}$   measures  the  ``semi-classicality'' of the system.

It is convenient to express  the action  in terms of dimensionless variables
  as follows  $x= x'/m_{\sigma}$ such that  $x'$ becomes a  dimensionless coordinate.  All distances now are measured in units of $m^{-1}_{\sigma}$. With this rescaling   the action   (\ref{thetaaction}) assumes  a very nice form:
\be	\label{action}
	S &=& {\cal N} \int_{\mathbb{R}^{3}} d^{3}x \sum_{n=1}^{N}\frac{1}{2}\left( \nabla \sigma_n \right)^{2}
	      \nonumber \\
	  &-& {\cal N} \int_{\mathbb{R}^{3}}  d^{3}x \sum_{a = 1}^{N} \cos \left(\sigma_{n} - \sigma_{n+1} 
	      + \frac{\theta}{N} \right), 
\ee 
with $\sigma_{N+1}$ identified with $\sigma_{1}$.  In formula (\ref{action}) we used $x$ as the dimensionless coordinate (rather than $x'$) to simplify notations.  
The Lagrangian entering the action (\ref{action}) is then dimensionless  with a large semiclassical prefactor ${\cal N}\gg 1$ defined by (\ref{debye}).

\section{Classification Scheme of the Vacuum States}\label{classification}
 We start with a short  overview of a well-known formal mathematical analogy between the construction of the $|\theta\ra$ vacuum states in gauge theories and Bloch's construction of the allowed/forbidden bands  in CM physics (see e.g. \cite{Shifman:2012zz}). 
The large gauge transformation operator $\cal{T}$ plays the role of the crystal translation operator in CM physics. $\cal{T}$ commutes with the Hamiltonian $H$ and changes the topological sector of the system 
 \be
 \label{large_gauge_transform}
 {\cal{T}}|m\ra=|m+1\ra, ~~ [H, {\cal{T}}]=0, 
 \ee
 such that 
   the $|\theta\ra $-vacuum state  is an eigenstate of the large gauge transformation operator $\cal{T}$:
\be
|\theta\rangle =\sum_{m\in\mathbb{Z}} e^{im \theta} |m\rangle, ~~~  {\cal{T}} |\theta\ra=e^{-i\theta}  |\theta\ra .\nonumber
\ee
The $\theta$ parameter in this construction plays the role of  the ``quasi-momentum" $\theta\rightarrow qa$ of a quasiparticle propagating in the allowed energy band in a crystal lattice with unit cell length $a$. 

An important element, which is typically skipped in presenting  this   analogy but which  plays a key role in our studies is the presence of the  Brillouin zones classified by integers $k$. Complete classification can be either presented in the so-called extended zone scheme where $-\infty < qa < +\infty$, or the reduced  zone scheme where each state is classified by two numbers, the quasi-momentum $-\pi \leq qa \leq +\pi$ and the Brillouin zone number $k$. 
 
In  the classification of the vacuum states, this corresponds to describing the system by two numbers $|\theta, k\ra$, where $\theta$  is assumed to be varied in the conventional range $\theta\in [0, 2\pi)$, while the integer $k$ describes the ground state (for $k=0$) or the excited metastable vacuum states ($k\neq 0$).  In most studies devoted to the analysis of the $\theta$ vacua, the questions  related to the metastable vacuum states have not been addressed. Nevertheless, it has been known for some time  that  the   metastable vacuum states must be present in non-abelian  gauge  systems in the large $N$ limit  \cite{Witten:1980sp}.
A similar conclusion  also  follows from the  holographic description of QCD as originally discussed in \cite{Witten:1998uka}. 

 In the present context  the metastable  vacuum states have been  explicitly constructed in a weakly coupled ``deformed QCD" model  \cite{Bhoonah:2014gpa}. 
We follow this construction by keeping both: the metastable states as well as $\theta\neq 0$ states, such that our complete classification is $|\theta, m\ra$
when the integer $m$ describes the metastable states for a given $\theta\in [0, 2\pi)$. In terms of the CM physics we use the so-called reduced zone scheme, rather than the extended zone scheme as defined above. 
 
The Euclidean potential density for the $\sigma$ fields assumes the following form (\ref{action}) 
\be	
\label{potential}
	U(\bm{\sigma}, \theta) = {\cal{N}} \sum_{n=1}^{N} \left[ 1 - \cos \left( \sigma_{n} - \sigma_{n+1}+\frac{\theta}{N} \right) \right],
\ee
where  we have added a constant  term  so that the potential is positive semi-definite. In eq. (\ref{potential}) 
the field  $\sigma_{N+1}$ is identified with $\sigma_{1}$ as before.  

The lowest energy state,   is the state with all $\sigma$ fields sitting at the same value ($\sigma_n = \sigma_{n+1}$) and has zero energy.
This is clearly the true ground state of the system, but there are also potentially some higher energy metastable states even for $\theta=0$.
For an extremal state we must have
\be
	\frac{\partial U}{\partial \sigma_n} = 0
\ee
for all $n$, which gives immediately
\be	\label{necessary}
	\sin \left( \sigma_{n} - \sigma_{n+1} +\frac{\theta}{N} \right) = \sin \left( \sigma_{n-1} - \sigma_{n} +\frac{\theta}{N} \right).~~
\ee
A necessary condition for a higher energy minimum of the potential is thus that the $\sigma$ fields are evenly spaced around the unit circle or (up to a total rotation),
\be	\label{sigmanecessary}
	\sigma_n = m \frac{2 \pi n}{N},
\ee
where $m$ is an integer which labels  the metastable states in the extended classification scheme $|\theta, m\ra$. This parameter plays the same role as the  
the Brillouin zone number $k$ in CM physics as discussed above.
A sufficient condition is then
\be
\label{derivative}
	\frac{\partial^2 U}{\partial \sigma_n^2} > 0,
\ee
again for all $n$. This gives us
\be
	\cos \left( \sigma_{n} - \sigma_{n+1} +\frac{\theta}{N} \right) + \cos \left( \sigma_{n-1} - \sigma_{n} +\frac{\theta}{N} \right) > 0, ~~~~~
\ee
which using (\ref{sigmanecessary}) gives
\be	\label{sufficient}
	\cos \left(  \frac{2 \pi m}{N} -\frac{\theta}{N}  \right) > 0.
\ee
This condition determines possible metastable states $m$ for a given $\theta\in [0, 2\pi)$ and $N$. 
From (\ref{sufficient}) it is quite obvious that metastable states always  exist for sufficiently large $N$ even for $\theta=0$, which is is definitely consistent with old and very generic arguments \cite{Witten:1980sp}. In our simplified version of the theory one can explicitly see how these metastable states emerge in the system, and how they are classified
in terms of the scalar magnetic potential fields  $\bm{\sigma} (\mathbf{x})$ for arbitrary $\theta$.

One should   remark here that a non-trivial solution for $\theta=0$ with $m\neq 0$ in (\ref{sufficient}) does not exist  in the ``deformed QCD"  model for the lowest $N = 2, 3, 4$ as it was originally discussed in \cite{Bhoonah:2014gpa}.   However, for sufficiently large  $\theta\neq 0$ the metastable states always emerge  for $N\geq 3$, while $N=2$, as usual, requires a special treatment \cite{Unsal:2012zj}. What is more important is that eq. (\ref{sufficient}) 
explicitly shows that at $\theta=\pi$   a metastable state with $m=1$ becomes degenerate with the ground state with $m=0$ and the level crossing phenomenon  takes place precisely as it was originally described in \cite{Thomas:2011ee} for this specific model. When $\theta$ further increases the metastable state becomes the  lowest energy state of the system (\ref{potential}) for the given $\theta$.

\section{Oblique Confinement for $|\theta, m\ra$ states}\label{oblique}
To understand the physical meaning of the solutions describing the nontrivial metastable vacuum states, one should compute the vacuum expectation value 
$\langle {\cal{M}}_a (\mathbf{x})\rangle$ of  the magnetization for a given   state $|\theta, m\ra$ classified by two  parameters $m, \theta$ as presented in previous section \ref{classification}.
The corresponding  operator  ${\cal{M}}_a (\mathbf{x})$ is defined as the creation operator of a single monopole of type $\alpha_{a}$ at point $\mathbf{x}$. It  has been originally computed for the ``deformed QCD"  model in  \cite{Thomas:2011ee}.
The corresponding computations can be easily generalized for arbitrary $\theta\neq 0$.  
The result of computations is
\be
\label{operator}
{\cal{M}}_a (\mathbf{x}) =e^{i \left(\alpha_{a} \cdot \bm{\sigma} (\mathbf{x})+\frac{\theta}{N}\right)}.
\ee
In the computation of (\ref{operator}) it has been assumed that the external magnetic source is infinitely heavy. If one identifies the corresponding magnetic source
with the monopoles from the ensemble  then the corresponding operator is accompanied by conventional classical contribution $8\pi^2/(g^2N)$.
Therefore, the resulting creation operator of a single monopole of type $\alpha_{a}$ at point $\mathbf{x}$ assumes the form
\be
\label{operator1}
{\cal{M}}_a (\mathbf{x}) =e^{-\frac{8\pi^2}{g^2N}} \cdot e^{i \left(\alpha_{a} \cdot \bm{\sigma} (\mathbf{x})+\frac{\theta}{N}\right)}.
\ee
This expression for the operator identically coincides for $N=2$ with formula (\ref{magnetic-operator}) derived in drastically different way
by starting from ${\cal{N}}=2$ supersymmetric model and 
  breaking the supersymmetry.

Now we are in position to compute  the vacuum expectation value $\la \theta, m|{\cal{M}}_a (\mathbf{x})|\theta, m\ra  $ describing  the magnetization of the system. It  can be easily computed for each  given  state $|\theta, m\ra$.   Indeed, 
using the solutions  (\ref{sigmanecessary}),  
the magnetization assumes the form 
\be
\label{magnetization}
\la \theta, m|{\cal{M}}_a (\mathbf{x})|\theta, m\ra  \sim \exp{\left[{ i\frac{\theta}{N} -i\frac{2\pi m}{N}}\right]},
\ee 
where one should pick up a proper branch which satisfies condition (\ref{sufficient}) describing the lowest energy state.

A different, but equivalent way to describe  all these     $|\theta, m\ra$ states  is to compute the expectation values for the topological density operator for those states. By definition,
\be
\label{top}
&&\la \theta, m| \frac{1}{16 \pi^{2}} \mathrm{tr} \left[ F_{\mu\nu} \tilde{F}^{\mu\nu} \right]|\theta, m\ra \equiv-i\frac{\partial S_{\rm dual}(\theta)}{\partial \theta} \nonumber\\
&&= i\frac{\zeta}{L}\sin \left(\frac{2\pi m}{N}-\frac{\theta}{N}\right). 
\ee
where the dual action $S_{\rm dual}(\theta)$ is given by  (\ref{thetaaction}).  
The imaginary $i$ in this expression should not confuse the readers as we work in the Euclidean space-time.
In Minkowski space-time this expectation value is obviously a real number.
A similar phenomenon is known to occur in the exactly solvable two dimensional Schwinger model wherein the expectation value for the electric field in the Euclidean space-time has an $i$.  
The expectation value (\ref{top}) is the order parameter of a given $|\theta, m\ra$ state. 

As expected, the ground state with $m=0$ at $\theta=0$ 
the expectation value (\ref{top}) vanishes, which of course, implies that the ground state respects  $\cal{P}$ and  $\cal{CP}$   symmetries.
It is not the case for a generic states with $\theta\neq 0$. These symmetries are also broken    for metastable states $m\neq 0$ even for $\theta=0$ as emphasized in  \cite{Bhoonah:2014gpa}.

The fact that the confinement in this model is due to the condensation of fractionally charged monopoles has been known since the original paper \cite{Unsal:2008ch}. Our original claim here is that the microscopical structure of the arbitrary $|\theta, m\ra$ states 
  can be also thought of as a condensate of the {\it same fractionally charged monopoles}.
 The only difference in comparison with the original construction  \cite{Unsal:2008ch} is that   the corresponding  magnetization receives a non-trivial phase (\ref{magnetization}) which depends on $\theta$ and    integer number $m$ which plays the same role as $k$-th Brillouin zone   in the reduced   classification scheme in CM physics. 

Now we want to present a few additional arguments suggesting that the confinement in this system is indeed
generated by the same magnetic monopoles with no trace for any dyons in this system which would provide a  conventional  ``dyon mechanism". Indeed,  the presence of the electrically charged dyons would imply that the interaction pattern between two BPS dyons at distance $r$ must have the following structure
\be
\label{dyons}
\sim \frac{1}{4\pi r}\left[e^2 q\cdot q'+\left(\frac{2\pi}{e^2}\right) m\cdot m'\right].
 \ee
 At the same time there is no trace for such kind of interaction in the original partition function which 
 assumes the form  \cite{Unsal:2008ch} 
 \be
 \label{partition}
  e^{ \frac{-2 \pi^{2} L}{g^{2}}
		\left[\sum_{a,b = 1}^{N} \sum_{k = 1}^{M^{(a)}} \sum_{l = 0}^{M^{(b)}} \alpha_{a} \cdot \alpha_{b} \;
		Q_{k}^{(a)} Q_{l}^{(b)} G ( \mathbf{x}_{k}^{(a)} - \mathbf{x}_{l}^{(b)} ) \right]}, ~~~~
		\ee
		where $G ( \mathbf{x}_{k}^{(a)} - \mathbf{x}_{l}^{(b)} ) $ is the corresponding Green's function. Precisely this interaction generates  
 the dual  action   (\ref{thetaaction}) which provides a proper low energy description of the system.  One can explicitly see that there is no electric portion of the interaction in  formula (\ref{partition}), in contrast with the   anticipated structure   expressed    as  (\ref{dyons}) which   is the conventional formula describing  the interaction of two non-BPS dyons carrying simultaneously the magnetic and  electric charges. In Section \ref{weak-coupling}  we come back to this point re-emphasize it from a different perspective. 

The argument presented above obviously  dismisses the presence of the electric charge of the constituents. It also    evidently  rises the following question. How does the self-duality  work  in this  case if the electric charges are not carried by the constituents? The answer is as follows: the BPS self-duality for the monopole's solutions is perfectly satisfied. However, the electric portion of the self-duality  equation is due to the generation of the nontrivial holonomy rather than due to the electric charges of the dyons.  Indeed, the self- duality equations for the monopoles assume the conventional form 
\be
\label{selfduality}
		D_i A_4^a=B_i^a, ~~~
		\left< A_4^{(a)} \right> = \frac{2\pi}{N L}\mu^a, ~~~ \mu^a \cdot \alpha_{b} =  \delta^{a}_{b}, 
				\ee
		which is precisely the key element in the original construction \cite{Unsal:2008ch} when the holonomy $\left< A_4^{(a)} \right>$ plays the role of the vacuum expectation value for the Higgs field. 
		
Another related question can be formulated  as follows. The topological charge operator is normally expressed as the product  of the magnetic and electric fields,  $q(\mathbf{x}) \sim \mathbf{E}^{(a)} (\mathbf{x})\cdot \mathbf{B}^{(a)} (\mathbf{x})$. At the same  time we claim that only magnetic monopoles are present in the system. These monopoles   generate the oblique confinement, and saturate the vacuum expectation values  (\ref{magnetization}) and (\ref{top}). How does it work?
The answer is as follows. The topological charge operator assumes the form  
 		\be
		\label{magnetic-charges}
		\int_{\mathbb{R}^{3} \times S^{1}}  d^4x q(\mathbf{x})  &=&   
			    		   \int_{\mathbb{R}^{3} \times S^{1}} d^4x \frac{g }{4 \pi^2} \sum_{a = 1}^{N} \left< A_{4}^{(a)} \right>
				\left[ \nabla \cdot \mathbf{B}^{(a)} (\mathbf{x}) \right]\nonumber\\
				&=&  \int_{\mathbb{R}^{3}}  d^3x \frac{1}{N} \sum_{a = 1}^{N} \sum_{k = 1}^{M^{(a)}} Q_{k}^{(a)} \delta (\mathbf{r}_{k}^{(a)} - \mathbf{x}),
		\ee
where we integrated by parts and used  formula (\ref{selfduality}) 	for the holonomy $\left< A_4^{(a)} \right>$. One can explicitly see from (\ref{magnetic-charges})		that the only constituents of the system are fractionally charged magnetic monopoles located  at $ \delta (\mathbf{r}_{k}^{(a)} - \mathbf{x})$ with zero electric charges as the corresponding sources are entirely determined by the divergence of the magnetic field $\left[ \nabla \cdot \mathbf{B}^{(a)} (\mathbf{x}) \right]$. In other words, there is no trace for the  dyons to play any role  in the system.\footnote{The electric field symoble, ${\bm E^{(a)}}$, that appears in the topological charge operator does not represent a genuine electric field irrespective of using the symbol ${\bm E}$. In fact, in the BPS limit (e.g. in ${\cal N}=1$ supr Yang-Mills) this field mediates attractive force between similar charges, i.e., it plays the role of a dilaton or scalar field. As we break SUSY and go to the deformed pure Yang-Mills limit, the scalar field is gapped and we are left only with interactions due to the magnetic field, as given by Equation (\ref{partition}), i.e., there are only objects that carry magnetic charges. Hence, there are no dyons. The same conclusion will be reached in Section \ref{dyons-dynamics}.} 

Nevertheless, the gap is generated at $\theta\neq 0$, the confinement takes place  in the conventional manner through the condensation of the monopoles (\ref{magnetization}), 
the $\theta$ parameter enters all the observables precisely as it should. This example explicitly shows that the conventional view  that the confinement in gauge theories at $\theta\neq 0$ 	is a result of the condensation of the dyons cannot be correct, at least in this simplified ``deformed QCD" model.
 Furthermore, as the transition between the ``deformed QCD" model and strongly coupled gauge theories should be smooth, we expect that the picture presented above must hold in strongly coupled regime as well. These results should be contrasted with the  common lore which assumes that the oblique confinement at $\theta\neq 0$ is a result of condensation of the electrically charged dyons. In the next sections we consider supersymmetric models  to  understand the nature of this difference.

\section{Dyons and monopoles in ${\cal N}=2$ super Yang-Mills}\label{dyons-static}

Dyons and monopoles are the main nonperturbative players in confinement in mass deformed ${\cal N}=2$ super Yang-Mills on $\mathbb R^4$, as the monopole's condensation leads to the confinement of electric charge probes  at $\theta=0$ as originally discussed in \cite{Seiberg:1994rs}.
It is commonly assumed that these monopoles at $\theta\neq 0$ become the dyons as  a result of the Witten's effect \cite{Witten:1979ey}. The condensation of the dyons would lead to   the oblique confinement speculated long ago by t' Hooft \cite{tHooft:1981bkw}.   On the other hand, it is the pure monopoles and not the dyons  that lead to confinement in deformed Yang-Mills on $\mathbb R^3\times \mathbb S^1$ as explained above. To elucidate this difference and track what really happens as we go from ${\cal N}=2$ super Yang-Mills to deformed Yang-Mills on a circle, we start by reviewing the field contents of the former theory. We warn the reader that unlike in previous sections, now we care to distinguish between Minkowskian and Euclidean quantities. This is important to arrive at distinct conclusions. 

${\cal N}=2$ super Yang-Mills theory has a massless ${\cal N}=2$  hypermultiplet that contains four bosonic and four fermionic degrees of freedom, see e.g. the textbook \cite{Shifman:2012zz}. Alternatively, one can decompose the  ${\cal N}=2$ multiplet into two ${\cal N}=1$ mutiplets: a vector and chiral mutiplets, both are in the adjoint representation of the gauge group. The bosonic part of the Lagrangian of ${\cal N}=2$ super Yang-Mills is given by (in Minkowski space, where we work with the signature $\eta_{MN}=(+1,-1,-1,-1)$)
\begin{eqnarray}
\nonumber
{\cal L}=\frac{1}{g^2}\mbox{tr}\left[-\frac{1}{2}F_{MN}F^{MN}+D_M\phi^\dagger D^M \phi+\frac{1}{2}\left[\phi^\dagger\,,\phi\right]^2 \right]\,,\\
\label{N 2 lagrangian}
\end{eqnarray} 
where $M,N=0,1,2,3$, the covariant derivative is $D_M= \partial_\mu+i \left[A_M,\quad\right]$,  the field strength is $F_{MN}=\partial_M A_N-\partial_N A_M+i\left[A_M,A_N\right]$, and the Lie algebra generators $t^a$ are normalized as $\mbox{tr}\left[t^at^b\right]=\frac{\delta_{ab}}{2}$. For simplicity, we will mainly work with $SU(2)$ gauge group. Without loss of generality we can always choose the vev of $\phi$  to be along the Cartan generators, i.e. along the $t^3$ direction in the $SU(2)$ case. Therefore, we take $\phi=vt^3$ such that $SU(2)$ is broken down to $U(1)$, i.e., we are in the Coulomb branch, and the potential term $\mbox{tr}\left[\phi^\dagger\,,\phi\right]^2$ vanishes, i.e., we are in the Bogomol'nyi-Prasad-Sommerfield (BPS) limit. we can also take the gauge invariant field $u=\mbox{tr}\left[\phi^2\right]$ to parametrize the moduli space of the gauge theory. The theory has a strong coupling scale $\Lambda$ such that in the limit $u\gg \Lambda^2$ the theory is in the weakly coupled regime\footnote{To one-loop order we have $\frac{4\pi}{g^2}=\frac{2}{\pi}\log \frac{v}{\Lambda}$.},  $g \ll 1$.   


In the weakly coupled regime, both perturbative and nonperturbative spectra of the theory can be determined using semi-classical analysis. As we mentioned above, the theory is Higgsed down to $U(1)$, and therefore, the bosonic part of the perturbative spectrum consists of a massless photon and W-bosons of charges $\pm 1$ with respect to the unbroken $U(1)$. Since we are in the BPS limit, the nonperturbative spectrum can be obtained via the Bogomol'nyi completion of the energy functional (see \cite{Weinberg:2006rq} for a review):
\begin{eqnarray}
\nonumber
E&=&\frac{1}{g^2}\int d^3x~ \mbox{tr} \left[E_i^2+B_i^2+\left(D_0\phi\right)^2+\left(D_i\phi\right)^2\right]\\
\nonumber
&=&\frac{1}{g^2}\int d^3 x~\mbox{tr}\left[\left(B_i\mp \cos\alpha D_i\phi\right)^2+\left(E_i\mp \sin\alpha D_i\phi\right)^2\right.\\
\nonumber
&&\left.+\left(D_0\phi\right)^2\right]\pm v \cos\alpha  Q_M\pm v \sin\alpha  Q_E\\
&\geq& \pm v \cos\alpha Q_M\pm v\sin\alpha  Q_E\,,
\label{B completion}
\end{eqnarray}
where  $E_i=F_{i0}$, $B_i=\frac{1}{2}\epsilon_{ijk}F^{jk}$, and we have used integration by parts and the Bianchi identity $D_iB_i=0$\,. The magnetic and electric charges, $Q_M$ and $Q_E$,  are defined via 
\begin{eqnarray}
\nonumber
Q_M&=&\frac{2}{g^2v}\int d^2S_i \mbox{tr}\left[\phi B_i\right]\,,\\
Q_E&=&\frac{2}{g^2v}\int d^2S_i \mbox{tr}\left[\phi E_i\right]\,,
\end{eqnarray} 
where $S_i$ is a two-sphere at spatial infinity. Similarly, one can define the scalar (dilaton) charge as
\begin{eqnarray}
Q_S=\frac{2}{g^2v}\int d^2S_i \mbox{tr}\left[\phi D_i\phi\right]\,.
\end{eqnarray}
 The most stringent inequality $E\geq v\sqrt{Q_E^2+Q_M^2}$ is obtained by setting $\alpha=\tan^{-1}\left(\frac{Q_E}{Q_M}\right)$. The equality is  saturated by a configuration that satisfies the first order equations
\begin{eqnarray}
\nonumber
B_i&=&\pm\cos \alpha D_i\phi\,,\\
\nonumber
E_i&=&\pm\sin \alpha D_i \phi\,,\\
D_0\phi&=&0\,.
\label{first order equations}
\end{eqnarray} 

Equations (\ref{first order equations}), with the upper sign, are solved by the ansatz:
\begin{eqnarray}
\nonumber
A_i^a&=&\epsilon_{iam}\hat r^m\left[\frac{1-u(r)}{r}\right]\,,\\
A_0^a&=&\hat r^a j(r)\,,~~~
\phi^a=\hat r^a h(r)\,.
\label{solution label}
\end{eqnarray}
Substituting (\ref{solution label}) into (\ref{first order equations}) one finds the solution 
\begin{eqnarray}
\nonumber
u(r)&=&\frac{\tilde v r}{\sinh (\tilde v r)}\,,\\
\nonumber
h(r)&=&\frac{\sqrt{ Q_M^2+ Q_E^2}}{ Q_M}\left[\tilde v \coth\left(\tilde v r\right)-\frac{1}{r}\right]\,,\\
j(r)&=&-\frac{Q_E}{Q_M}\left[\tilde v  \coth\left(\tilde v r\right)-\frac{1}{r} \right]\,,
\label{solution functions}
\end{eqnarray}
where  $\tilde v=v\frac{Q_M}{\sqrt{Q_E^2+ Q_M^2}}$. Equations (\ref{solution functions}) constitute Julia-Zee dyon-particle \cite{Julia:1975ff}. This configuration has a total energy (mass)
\begin{eqnarray}
E=M=v\sqrt{Q_M^2+Q_E^2}\,.
\end{eqnarray}
 In the limit $r\rightarrow \infty$ one can use (\ref{solution label}) and (\ref{solution functions}) to show that
\begin{eqnarray}
E_i\sim\frac{Q_E x_i}{ r^3}\,,~~~
B_i\sim\frac{Q_M x_i}{ r^3}\,,~~~
D_i\phi\sim-\frac{Q_S x_i}{ r^3}\, , ~~~
\label{asymptotic fields}
\end{eqnarray}
where $Q_S=\frac{Q_M}{\cos\alpha}=\sqrt{Q_E^2+Q_M^2}$. Thus, the dyon mass satisfies the relation
 \be M=vQ_S\,.\ee
Using (\ref{solution functions}) in the energy functional (\ref{B completion}) we obtain the interaction energy of two BPS dyon-particles with charges $(Q_M,Q_E)$ and $(Q_M',Q_E')$ and located at $\bm r$ and $\bm r'$:
\begin{eqnarray}
E_{int}\sim g^2\frac{Q_EQ_E'+Q_MQ_M'-Q_SQ_S'}{\left|\bm r-\bm r'\right|}\,.
\label{int energy}
\end{eqnarray}
We see that, as expected, the interaction force of the $U(1)$ field is repulsive, while the dilaton field is attractive \cite{Bak:1997pc}.

 Dyons are genuine particles\footnote{More precisely, they are solitons.} that carry both electric and magnetic charges. Classically, a dyon can have an arbitrary electric charge, while it can only have quantized magnetic charge $Q_M=\frac{4\pi n}{g^2}$, where $n$ is a positive or negative integer, due to obvious topological reasons. However, quantum mechanical consistency demands that a pair of dyon-particles with charges $(Q_M, Q_E)$ and  $(Q_M',Q_E')$ must satisfy the Dirac quantization condition $Q_EQ_M'- Q_E'Q_M=n\frac{4\pi}{g^2}$, where $n$ is an integer. Therefore, both electric and magnetic charges must be quantized: $(Q_M,Q_E)=\left(\frac{4\pi n_M}{g^2},n_E\right)$, where $n_M,n_E \in \mathbb Z$, and we find that the BPS spectrum is given by
\begin{eqnarray}
M(n_M,n_E)=v\sqrt{n_M^2\left(\frac{4\pi}{g^2}\right)^2+n_E^2}\,.
\end{eqnarray}
The BPS masses $M(n_M,n_E)$ do not receive quantum corrections, thanks to the high level of supersymmetry in ${\cal N}=2$ super Yang-Mills. In addition, one can take into account the effect of the $\theta$-vacuum, $\frac{\theta}{32\pi^2} \tilde F_{MN}F^{MN}$, by  making the substitution $n_E\rightarrow n_E+n_M\frac{\theta}{2\pi}$, which is Witten's effect \cite{Witten:1979ey}. One finally finds:
\begin{eqnarray}
M(n_M,n_E, \theta)=v\sqrt{n_M^2\left(\frac{4\pi}{g^2}\right)^2+\left(n_E+n_M\frac{\theta}{2\pi}\right)^2}. ~~~~~
\label{BPS spectrum}
\end{eqnarray}
We could also set $n_E=0$, which is the limiting case of 't Hooft Polyakov monopole particles. However,  a single monopole has four collective coordinates: three translation coordinates and one coordinate corresponding to a $U(1)$ global transformation. The $U(1)$ collective coordinate is compact (remember that $U(1)$ is descendent from $SU(2)$, which is a compact group). The Hamiltonian corresponding to the compact coordinate is $H_{U(1)}=\frac{p_\phi^2}{2I}$, with $I=\frac{4\pi}{g^2 v}$.  Upon quantization, the magnetic monopole acquires an electric charge;  this is one of the eigenvalues of $H_{U(1)}$.  Thus, quantum fluctuations in the background of an 't Hooft Polyakov monopole dresses it with a quantized electric charge and gives rise to a dyon, with its mass given by the BPS expression (\ref{BPS spectrum}).  

While the spectrum (\ref{BPS spectrum}) is a well-established feature of ${\cal N}=2$ supersymmetry in weakly coupled regime at large $v$, the role of these dyons 
in confined strongly coupled regime is less understood. We review below some features of the system relevant for our studies by paying special attention 
to the dyon-particles. Precisely these degrees of freedom,   according to the conventional wisdom,  should condense at $\theta\neq 0$ and provide a precise realization for the oblique confinement as envisaged by 't Hooft \cite{tHooft:1981bkw}. 

As we approach the strong coupling regime of the theory, $v\sim \Lambda$, most of the dyon-particles decay except the ones with lowest charges $(1,0)$ or $(1,1)$, which become massless. This theory is electrically strongly coupled and magnetically weakly coupled. Therefore, the theory can be described by a dual ${\cal N}=2$ supersymmetric electrodynamic of massless monopoles or dyons. 

Now we insert a small mass term into the action with $m\ll \Lambda$. It breaks the symmetry from ${\cal N}=2$ down to ${\cal N}=1$. One could naively think that the oblique confinement might take place as a result of the dyon  condensation. However, the oblique confinement does not occur,
at least in weakly coupled regime \cite{DiPierro:1996zj,Konishi:1996iz}. The basic reason for that is that the ``pure monopoles" rather than  dyons condense at both points $u=\pm \Lambda^2$ as argued in \cite{DiPierro:1996zj,Konishi:1996iz}. This is in spite of the fact that near $u=-\Lambda^2$ the dyons $(1,1)$ rather than monopoles $(1,0)$ become massless  particles. The absence of the oblique confinement in the system is obviously consistent with our analysis of the ``deformed QCD" model in  section \ref{oblique}.  However, one cannot make a definite conclusion with a  large supersymmetry breaking in this construction when the question on oblique confinement remains open \cite{DiPierro:1996zj,Konishi:1996iz}. It should be contrasted with results of section \ref{oblique} where the transition to strongly coupled regime of ordinary QCD is expected to be smooth as argued in Section \ref{smooth}.  

One can also insert $N_f$  flavours into the system \cite{Konishi:1998mk}.  It turns out that the oblique confinement occurs for $N_f=3$ model, but does not occur for $N_f=2$ nor for $N_f=1$ models. All the arguments of refs \cite{DiPierro:1996zj,Konishi:1996iz,Konishi:1998mk} are crucially depend on the specific properties of supersymmetric theories. Therefore, it is not obvious if one can learn any lessons for ordinary QCD. With this motivation in mind we consider the Witten-Sieberg model being formulated on $\mathbb R^3\times \mathbb S^1$ when one can approach the weakly coupled regime by varying the size of $\mathbb S^1$.

\section{Dyon-instantons vs. Monopole-instantons on $\mathbb R^3\times \mathbb S^1$}\label{dyons-dynamics}

In this section   we show that the nonperturbative sector of ${\cal N}=2$ on $\mathbb R^3\times \mathbb S^1$ consist of a variety of dyons, similar to our previous discussions. However, in this section we consider the Euclidean, rather than Minkowski formulation. Therefore we compute the Euclidean action generated by the psudo- particles, rather than particles. To avoid confusion with terminology we coin  the corresponding pseudo-particles the dyon-instantons to emphasize on their Euclidean nature.  We will show that  for sufficiently small $\mathbb S^1$ circle in weakly coupled regime where our computations are under control, the partition function is saturated by  the tower of monopole-instantons rather than dyon-instantons. In our presentations we closely follow \cite{Chen:2010yr,Poppitz:2011wy,Chen:2011gk,Dorey:2000dt,Dorey:2000qc}.

\subsection{$\mathbb R^3\times \mathbb S^1$: the large circle limit}\label{strong coupling}

We start our treatment by compactifing the $x^3$-direction over a circle and considering the Euclidean version of the theory. The Euclidean time direction will be denoted by $x_4$ such that $x_4\equiv ix^0$, while the rest of coordinates are left intact. We also define the Euclidean fields $\hat A_i=-A^i$, and $\hat A_4=-iA_0$. Again, we assume that $v\gg \Lambda$, and hence, the theory is in its semi-classical regime. The Euclidean action is given by
\begin{eqnarray}
&&S_E=\frac{1}{g^2}\int_{\mathbb R^3 \times \mathbb S^1}\mbox{tr}\left[\frac{1}{2}\hat F_{MN}\hat F_{MN}+ \left(\hat D_M\phi\right)^2 \right]\\ 
&=&\frac{1}{g^2}\int_{\mathbb R^3 \times \mathbb S^1}\mbox{tr}\left[\left(\tilde E_\mu\right)^2+\left(\tilde B_\mu\right)^2+ \left(\hat D_3\phi\right)^2+\left(\hat D_\mu\phi\right)^2  \right]\,,\nonumber
\label{euclidean action}
\end{eqnarray}  
where $M,N=1,2,3,4$, $\hat F^a_{MN}=\partial_M \hat A^a_N-\partial_N\hat A^a_M+f^{abc}\hat A^b_M\hat A^c_N$, $\hat D_M\phi^a=\partial_M \phi^a+f^{abc}\hat A^b_M\phi^c$, where $f^{abc}$ are the group structure constants. We also defined $\tilde E_\mu=\hat F_{\mu3}$, $\tilde B_\mu=\frac{1}{2}\epsilon_{\mu\nu\alpha}\hat F_{\nu\alpha}$, where $\mu,\nu=1,2,4$. Notice that here we distinguish between the electric and magnetic fields $\hat E_i\,,\hat B_i$, where $i=1,2,3$, and $\tilde E_\mu$ and $\tilde B_\mu$. Although not mandatory, this distinction is convenient since it will enable us to keep track of various quantities. Comparing the Euclidean action (\ref{euclidean action}) with the energy functional (\ref{B completion}), we immediately reveal that a finite action solution can be obtained using the exact same procedure we followed to obtain Julia-Zee dyon-particles. The existence of a finite action solution demands that the fields profile are independent of $x_3$ (exactly like the dyon-particle solution is independent of $x^0$). One then can think of this solution as wrapping around the $x_3$-direction, and hence, in the Euclidean setup we obtain {\it dyon-instantons}\footnote{ One should not confuse 
these dyon-instantons with  the gauge configurations considered in refs. \cite{Liu:2015ufa,Liu:2015jsa,Larsen:2015vaa}, which were (incorrectly) coined as the dyons or dyon-instantons. Those configurations representing the instanton constituents from refs  \cite{Liu:2015ufa,Liu:2015jsa,Larsen:2015vaa} do not carry the electric charges and should be considered as 
monopoles-instantons  in our classification scheme.}  to be contrasted with {\it dyon-particles} considered in Section \ref{dyons-static}. Taking the length of the $\mathbb S^1$ circle to be $L$, we immediately find that the action of the BPS dyon-instanton is given by
\begin{eqnarray}
&&S(n_M,n_E, \theta)=LM(n_M,n_E, \theta)\\
&=&Lv\sqrt{n_M^2\left(\frac{4\pi}{g^2}\right)^2+\left(n_E+n_M\frac{\theta}{2\pi}\right)^2}\,.\nonumber
\label{BPS dyons}
\end{eqnarray}
In addition, two  BPS dyon-instantons carrying charges $(Q_M,Q_E)$ and $(Q_M',Q_E')$  and located at $\bm r$ and $\bm r'$ in the Euclidean space will interact as in (\ref{int energy}):
\begin{eqnarray}
S_{\rm int}\sim g^2\frac{Q_EQ_E'+Q_MQ_M'-Q_SQ_S'}{\left|\bm r-\bm r'\right|}\,,
\label{action int}
\end{eqnarray}
where  the Euclidean radial coordinate is defined as $r=\sqrt{x_1^2+x_2^2+x_4^2}$, such that  the profile functions in (\ref{solution label}) now depend on the newly defined $r$.  
Since dyon-instantons have a finite action, they will contribute to the Euclidean partition function. On $\mathbb R^3 \times \mathbb S^1$ the gauge potential $\hat A_3$ is a compact scalar with period $L$: $\hat A_3\cong \hat A_3+1/L$. For convenience let us define $\omega$ as
 \be\hat A_3\equiv \omega /L\,.\label{definition of omega}\ee
 In addition, we can go to a dual description such that 
\be\hat F_{\mu\nu}=\frac{g^2}{2\pi L}\epsilon_{\mu\nu\alpha}\partial_\alpha \sigma\,.\label{definition of sigma}\ee
 Again, one can show that $\sigma$ is a compact scalar with period $2\pi$. Let us also define $\Phi=\phi/L$. Then, in terms of the scalars $\omega\,,\sigma$, and $\Phi$,  the insertion of a dyon-instanton in the partition function can be represented by the vertex:
\begin{eqnarray}
{\cal D}&=&e^{-S(n_M,n_E, \theta)}\\
\nonumber
&\times& e^{i\left(n_E+\frac{\theta}{2\pi}n_M\right) \omega+in_M \sigma+\sqrt{n_M^2\left(\frac{4\pi}{g^2}\right)^2+\left(n_E+\frac{\theta}{2\pi}n_M\right)^2}\Phi}\,\\
&\times& \mbox{fermion zero modes}\,.
\end{eqnarray}
In the absence of fermion zero modes one can easily show that this vertex will also reproduce the interaction (\ref{action int}) such that $\omega$ mediates the force between $Q_E$ charged objects,  $\sigma$ mediates the force between $Q_M$ charged objects, while $\phi$ mediates the force between $Q_S$ charged objects\footnote{This can be shown by writing the abelian part of the kinetic term $F_{MN}F_{MN}+(D_M\phi)^2$ in terms of the three-dimensional fields $\omega,\sigma$, and $\Phi$:
\begin{eqnarray}
\mbox{K.E.}=\frac{1}{2g^2}\left(\partial_\mu \omega\right)^2+\frac{1}{2g^2}\left(\partial_\mu \Phi\right)^2+\frac{g^2}{8\pi^2}\left(\partial_\mu\sigma\right)^2
\end{eqnarray}
which is derived in Section (\ref{witten}). Then we insert the vertex ${\cal D}(x_1)$ in the partition function and solve for the quadratic Lagrangian to obtain expression (\ref{action int}).}. This interaction is repulsive for both $\omega$ and $\sigma$ fields (since both $\omega$ and $\sigma$ are parts of the electromagnetic $U(1)$ field; notice the imaginary number $i$ in front of these fields), while it is attractive for $\Phi$ (a scalar field; notice the absence of $i$ in front of it). 

A key point of this subsection is that the dyons are generic configurations of the system. The interaction pattern (\ref{action int}) obviously shows that 
they would be  genuine static dyons if one treats the Euclidean $x_3$ coordinate as a time variable.  Based on these   configurations one could naively think that oblique confinement should occur as a result of the condensation of the dyons as the generic gauge configurations of the system. Nevertheless, as we demonstrate  in next subsection \ref{weak-coupling}  this naive picture is incorrect: if one proceed with computations in theoretically controllable way,  the confinement occurs as a result of the monopole's (not   dyon's) condensation for arbitrary $\theta\neq 0$, similar to our analysis in Section \ref{oblique} in ``deformed QCD" model.

\subsection{ $\mathbb R^3\times \mathbb S^1$: the small circle limit}\label{weak-coupling}
Our goal now is to consider the small circle limit where computations can be carried out in theoretically controllable way. With this goal in mind 
 we ignore the fermion zero modes\footnote{Fermion zero modes in the duality we consider below is a subtle issue that yet to be understood, see \cite{Poppitz:2011wy,Chen:2010yr}.} and consider a tower of dyon-instantons with a unit magnetic charge and an arbitrary number of electric charges $(Q_M,Q_E)=(\frac{4\pi}{g^2},n_E)$:
\begin{eqnarray}
\nonumber
{\cal S}&=&\sum_{n_E \in \mathbb Z} e^{-S(1,n_E, \theta)}\\
&&\times e^{i\left(n_E+\frac{\theta}{2\pi}\right) \omega+i \sigma+\sqrt{\left(\frac{4\pi}{g^2}\right)^2+\left(n_E+\frac{\theta}{2\pi}\right)^2}\Phi}\,.
\label{tower sum}
\end{eqnarray}
Then, one can approximate the sum in (\ref{tower sum}) as
\begin{eqnarray}
\nonumber
{\cal S}&\cong&  e^{-\frac{4\pi (Lv-\Phi)}{g^2}+i\sigma}\\
&\times& \sum_{n_E \in \mathbb Z} e^{i\left(n_E+\frac{\theta}{2\pi}\right) \omega-\frac{g^2 Lv}{8\pi}\left(n_E+\frac{\theta}{2\pi}\right)^2}\,.
\label{approx tower sum}
\end{eqnarray}
 A fast convergence of the series demands that $Lv\gg \frac{4\pi}{g^2}\gg 1$. Therefore, for a very large circle the sum will rapidly converge. For a small circle, however, the series is poorly convergent and a method of resummation is indispensable. To achieve this, we use the Poisson resummation formula defined as:
\begin{eqnarray}
\nonumber
\sum_{n_E \in \mathbb Z} f(n_E)&=&\sum_{n_W \in \mathbb Z} \tilde f(n_W)\,,\\
\tilde f(n_W)&=&\int dk f(k)e^{-2\pi n_W k}\,.
\end{eqnarray}
Applying this method to the series (\ref{tower sum}) we find \cite{Chen:2010yr}, modulo pre-exponential factor,
\begin{eqnarray}
S\cong \sum_{n_W \in \mathbb Z} e^{i\sigma-\frac{4\pi}{g^2}\sqrt{(Lv-\Phi)^2+\left(\omega+2\pi n_W\right)^2}+in_W\theta}\,.
\label{exact monopole sums}
\end{eqnarray}
In the limit $vL\gg \omega$ we obtain the approximation
\begin{eqnarray}
S\cong e^{-\frac{4\pi (Lv-\Phi)}{g^2}+i\sigma} \sum_{n_W \in \mathbb Z}e^{-\frac{2\pi }{g^2 Lv}\left(\omega+2\pi n_W\right)^2+in_W\theta}\,.
\label{monopole sums}
\end{eqnarray}
The series (\ref{monopole sums}) is rapidly convergent\footnote{Remember that we are still in the semi-classical regime $g\ll1$. Thus, the series (\ref{monopole sums}) is valid in the parameter range $\frac{4\pi}{g^2}\gg Lv\gg \omega$.} in the small $\mathbb S^1$ limit $vL\ll1$. A careful inspection of (\ref{exact monopole sums}) reveals that the sum is over a tower of twisted monopole-instantons that carry magnetic charges $\pm1$ and winding numbers $n_W \in \mathbb Z$, as we show in details at the end of this section. This is a remarkable result since in the small circle limit we can think only about monopole-instantons instead of dyon-instantons. This claim holds for any $\theta\neq 0$,   as is evident from (\ref{exact monopole sums}).    

Let us now make the shift $\omega\rightarrow \omega+\Omega$, where $0<\Omega<2\pi$ is a background holonomy (remember that $\omega$ is the scaled $\hat A_3$ component), in (\ref{exact monopole sums}). It will suffice to consider only the two terms $n_W=0$ and $n_W=-1$. For small fluctuations of $\Phi$ and $\omega$ the terms $n_W=0$ and $n_W=-1$ are given by
\begin{eqnarray}
\nonumber
{\cal M}_{0}&=&e^{-S_{0}}e^{i\sigma+\frac{4\pi}{g^2}\frac{\Omega\left(\omega-\frac{Lv}{\Omega}\Phi\right)}{\sqrt{(Lv)^2+\Omega^2}}}\,,\\
{\cal M}_{-1}&=&e^{-S_{-1}}e^{i\sigma-i\theta+\frac{4\pi}{g^2}\frac{\left(\Omega-2\pi\right)\left(\omega-\frac{Lv}{\Omega-2\pi}\Phi\right)}{\sqrt{(Lv)^2+\left(\Omega-2\pi\right)^2}}}\,,
\label{both monopoles}
\end{eqnarray}
where $S_{0}=\frac{4\pi\sqrt{(Lv)^2+\Omega^2}}{g^2}$ and $S_{-1}=\frac{4\pi\sqrt{(Lv)^2+\left(\Omega-2\pi\right)^2}}{g^2}$ are respectively the actions of BPS and twisted (or Kaluza-Klein)  monopole-instantons. Notice that both monopole-instantons have positive magnetic charges, $n_M=1$, as is evident from the same sign in front of $i\sigma$ in (\ref{both monopoles}). This should be expected since the series (\ref{monopole sums}) originated from the sum over a tower of dyon-instantons all having the same magnetic charge $n_M=1$. Also, the imaginary number in front of $\sigma$ means that objects carrying the same magnetic charges will experience a repulsive force, which is also expected. The interesting thing, though, is the absence of $i$ in front of $\omega$ and $\Phi$, which means that the combination of the fields $\omega-\frac{Lv}{\Omega}\Phi$ or $\omega-\frac{Lv}{\Omega-2\pi}\Phi$ mediates a scalar force rather than an electromagnetic one. This is a fascinating phenomenon since we start with a tower of dyon-instantons at large $\mathbb S^1$.  The dyon-instantons experience  a repulsive electromagnetic force (for both electric $\omega$ and magnetic $\sigma$ parts) as in (\ref{action int}), in addition to a scalar force (mediated by the scalar field $\Phi$). Then, we resum over the electric charges of this tower, using Poisson resummation formula, to find that at a small $\mathbb S^1$ the electric force is incarnated as a scalar force. 
Using $Q_b$  to denote the charge of the monopole-instanton under any of these combinations, we find that two monopoles carrying charges $(Q_M,Q_b)$ and $(Q_M',Q_b')$ and located at  $\bm r$  and $\bm r'$ experience a force
\begin{eqnarray}
S_{int}\sim\frac{Q_MQ_M'-Q_bQ_b'}{\left|\bm r-\bm r'\right|}\,.
\label{monopole forces}
\end{eqnarray}
In particular, for the BPS and twisted monopole-instantons we have $|Q_M|=|Q_b|=\frac{4\pi}{g^2}$. 
This formula obviously shows that the only configurations which contribute to the partition function in the regime, where computations are under complete theoretical control, are the monopoles and twisted monopoles, but not the dyons carrying the electric charges.  

The last element  which completes our analysis of this subsection is the demonstration that  
  the sum in (\ref{monopole sums}) is indeed over a tower of twisted monopole-instantons. To this end, we set $\hat D_3\phi=0$ in (\ref{euclidean action})  (notice that this is exactly compatible with the third equation in (\ref{first order equations})) and we also set $\partial_3=0$. This at least is enough to obtain the zero-winding number (BPS) monopole-instanton. Monopoles with higher winding numbers (twisted monopoles) can be obtained by replacing $\hat A_3\rightarrow \hat A_3+\frac{2\pi n}{L}$. The lowest (zero-winding) monopole-instanton action reads
\begin{eqnarray}
\nonumber
S_E&=&\frac{2}{g^2}\int_{\mathbb R^3 \times \mathbb S^1}\mbox{tr}\left[\left(\hat D_\mu A_3\right)^2+\left(\hat B_\mu\right)^2+\left(\hat D_\mu \phi\right)^2\right]\\
\nonumber
&=&\frac{2}{g^2}\int\mbox{tr}\left[\left(\hat D_\mu\hat A_3\mp \sin\beta \tilde B_\mu\right)^2+\left(\hat D_\mu\phi\mp \sin\beta \tilde B_\mu\right)^2\right.\\
\nonumber
&&\left. \pm 2\sin\beta \hat D_\mu\hat A_3\tilde B_\mu\pm 2\cos\beta \hat D_\mu\phi \tilde B_\mu\right]\\
&\geq&L Q_M\left[\pm 2 \frac{\Omega}{L} \sin\beta  \pm 2 v\cos \beta \right]\,,
\end{eqnarray}
where $v$ and $\frac{\Omega}{L}$ are respectively the vevs of $\phi$ and $\hat A_3$ and the vevs are taken along the fourth direction in color space\footnote{Remember that we are in a Euclidean setup where our infinite dimensions are taken along $x_1,x_2,x_4$. Given our numbering convention, then we also take the color space index $a$ to run over $1,2,4$, where the diagonal Pauli matrix is taken along the $4$-direction.}. We also defined  the magnetic charge: $Q_M=\frac{1}{g^2}\int dS_\mu \hat B_\mu^4$, where the integral is over a two-sphere at infinity.  The most stringent inequality  $S_E\leq  Q_M\sqrt{L^2v^2+\Omega^2}$ is obtained by setting $\tan\beta=\frac{\Omega}{Lv}$, while the inequality is saturated by 
\begin{eqnarray}
\nonumber
\hat D_\mu \phi&=&\pm \hat B_\mu \cos\beta\,,\\
\hat D_\mu A_3&=&\pm \hat B_\mu\sin \beta\,.
\label{mixed monopole equations}
\end{eqnarray} 
A linear superposition of  (\ref{mixed monopole equations}) can be written as
\begin{eqnarray}
\nonumber
\hat B_\mu&=&\pm\hat D_\mu \Psi_1\,,\\
0&=&\hat D_\mu \Psi_2\,,
\label{anti self dual eq}
\end{eqnarray}
where 
\begin{eqnarray}
\nonumber
\hat\Psi_1&=&\sin\beta \hat A_3+\cos\beta \phi\,\\
\hat\Psi_2&=&\cos\beta \hat A_3-\sin\beta \phi.
\label{definitions of psi}
\end{eqnarray}
Equation (\ref{anti self dual eq}) is the (anti)self-dual BPS monopole-instanton equation. The solution of the self-dual equation is
\begin{eqnarray}
\nonumber
\hat A_\mu^a&=&\epsilon_{\mu a\nu}\hat x^\nu\left[\frac{1-u(r)}{r}\right]\,,\\
\nonumber
\hat \Psi_1^a&=&\hat x h(r)\,,\\
\hat \Psi_2&=&0\,,
\label{monopole solution}
\end{eqnarray}   
where $r=\sqrt{x_1^2+x_2^2+x_4^2}$ and
\begin{eqnarray}
\nonumber
u(r)&=&\frac{\tilde v r}{\sinh (\tilde v r)}\,,\\
h(r)&=&\tilde v \left[\cosh(\tilde vr)-\frac{1}{r}\right]\,,
\label{profile solutions for monopoles}
\end{eqnarray}
and $\tilde v=\sqrt{\frac{\Omega^2}{L^2}+v^2}$\,. 

 Now, to obtain the twisted-monopole solutions we just need to make the substitution $\Omega\rightarrow \Omega+2\pi n_W$ for all integers $n_W$. Thus, the action of the twisted monopole-instantons with magnetic charge $\frac{4\pi}{g^2}$ is given by
\begin{eqnarray}
S_{n_W}=\frac{4\pi}{g^2}\sqrt{L^2v^2+\left(\Omega+2\pi n_W\right)^2}\,,
\end{eqnarray}
which is exactly the action in the sum (\ref{exact monopole sums}).

The main lesson to be learnt from these computations is as follows. The generic gauge configurations of the system obviously include the dyons. 
However, if one tries to compute the partition function in a theoretically controllable  region,    the corresponding configurations can be described exclusively in terms of the monopoles, without any trace of the dyons. It is perfectly consistent with our analysis of the ``deformed QCD" model in Section \ref{oblique},
where confinement is generated for any $\theta\neq 0$ as a result of condensation of the monopoles. In next subsection \ref{SUSY-breaking} we show that the picture also holds when supersymmetry is broken.

\subsection{Supersymmetry breaking}\label{SUSY-breaking}

 In order to break supersymmetry in a controlled way we first add a suitable scalar mass term $m_\phi$ for the field $\phi$ and  its super partner. In the limit $m_\phi\gg \Lambda$ the scalar decouples,  which in turn  breaks ${\cal N}=2$ down to ${\cal N}=1$. If this decoupling happens in the large $\mathbb S^1$ limit, then the theory flows to strong coupling regime, we loose theoretical control, and the dyon-instantons picture is no longer trusted. However, if the decoupling happens at a small $\mathbb S^1$, then the theory stays in its weakly coupled regime and preserves its center symmetry, i.e., $\Omega=\pi$. Setting $v=0$ (since the scalar $\phi$ decouples), defining $b=\frac{4\pi}{g^2}\omega$, and shifting $\sigma\rightarrow \sigma+\frac{\theta}{2}$, we find that the monopole-instanton operators (\ref{both monopoles}) are given by
\begin{eqnarray}  
{\cal M}_{0,1}= e^{-S_m}e^{ i\sigma \pm (b+i\frac{\theta}{2})}\,,
\end{eqnarray}
and $S_m=\frac{4\pi^2}{g^2}$.

In order to further break ${\cal N}=1$ we give the gaugino a mass larger than the strong scale. Again, we can guarantee that the theory is in the weakly coupled regime as long as the circle is kept sufficiently small. Preserving the center symmetry, however, requires that we add a double trace deformation. This theory is our ``deformed QCD" model though in this case it represents pure gauge Yang-Mills fields, see footnote \ref{terminology} in Introduction regarding this terminological convention.  In this case the scalar field $b$ is gapped and we end with the monopole operators:
\begin{eqnarray}  
\label{magnetic-operator}
{\cal M}_{0,1}= e^{-S_m}e^{i\sigma \pm i\frac{\theta}{2}}\,.
\end{eqnarray}
This expression identically coincides with formula (\ref{operator1}) for ``deformed QCD" model derived in drastically different way. 

We conclude this section with the following generic comment. In all cases when the computations can be performed in a theoretically controllable way,
the gauge configurations which saturate the partition function are the monopole-instantons for arbitrary $\theta\neq 0$.  This  claim holds for ${\cal N}=2$, ${\cal N}=1$, and non- supersymmetric ``deformed QCD" model.
This result  should be contrasted with conventional wisdom that the oblique confinement in the system for $\theta\neq 0$ 
occurs as a result of the condensation of the electrically charged  dyons.

\section{Witten's effect}
\label{witten}

Since there is no trace of dyons in the spectrum of theories on $\mathbb R^3 \times \mathbb S^1$ in the small circle limit, one may wonder how Witten's effect is realized in this case. The answer is that this effect can be seen for static  (non dynamical) electric or magnetic charges, i.e. Wilson or 't Hooft loops, which we use to probe the system \cite{Anber:2015wha}. In order to show this explicitly, we start from the abelian action written in Minkowski space:
\begin{eqnarray}
\nonumber
S&=&\int_{\mathbb R^3 \times \mathbb S^1} \frac{1}{4g^2}F_{MN}F^{MN}+\frac{\theta}{32 \pi^2}F_{MN}\tilde F_{MN}\\
&=& \int_{\mathbb R^3 \times \mathbb S^1} \frac{1}{2g^2}\left(E_\mu E_\mu-B_\mu B_\mu\right)-\frac{\theta}{8\pi^2}E_\mu B_\mu\,,
\label{total Min action}
\end{eqnarray}
where $\tilde F_{MN}=\epsilon_{MNPQ}F^{PQ}/2$. Next, we dimensionally reduce the action (\ref{total Min action}) by neglecting all dependence on the $x^3$-direction and  use the fields $\omega$ and $\sigma$  defined via (\ref{definition of omega}) and (\ref{definition of sigma}) (now in Minkowski space) to find\footnote{The duality relation  (\ref{definition of sigma}) can be incorporated into the action (\ref{total Min action}) using the auxiliary action $\frac{1}{4\pi} \int d^3 x \epsilon_{\mu\nu\alpha}\partial_\mu \sigma F_{\nu\alpha}$.}
\begin{eqnarray}
F^{\nu\rho}=\frac{g^2}{2\pi L}\epsilon^{\mu\nu\rho} \left(\partial_\mu\sigma+\frac{\theta}{2\pi}\partial_\mu \omega\right)\,,
\label{duality relation}
\end{eqnarray}
and
\begin{eqnarray}
\nonumber
S=-\frac{1}{2L}\int d^3 x \frac{1}{g^2}\left(\partial_\mu \omega\right)^2+\frac{g^2}{4\pi^2}\left(\partial_\mu\sigma+\frac{\theta}{2\pi}\partial_\mu \omega\right)^2\,.\\
\end{eqnarray}
From (\ref{duality relation}) we easily find (keeping in mind that the Greek letters run over $0,1,2$, while the Latin letters $M,N$ run over $0,1,2,3$)
\begin{eqnarray}
\nonumber
B_1&=&\frac{\partial_2 \omega}{L}\,,\quad B_2=-\frac{\partial_1 \omega}{L}\,,\\
\nonumber
B_3&=&\frac{g^2}{2\pi L}\left(\partial_t \sigma+\frac{\theta}{2\pi}\partial_t\omega\right)\,,\\
\nonumber
E_1&=&-\frac{g^2}{2\pi L}\left(\partial_2 \sigma+\frac{\theta}{2\pi}\partial_2 \omega\right)\,,\\
\nonumber
E_2&=&\frac{g^2}{2\pi L}\left(\partial_1 \sigma+\frac{\theta}{2\pi}\partial_1 \omega\right)\,,\\
E_3&=&-\frac{\partial_t \omega}{L}\,.
\label{electric and magnetic fields}
\end{eqnarray} 

A Wilson loop operator that measures the magnetic flux in the $y-z$ plane and warps around the $\mathbb S^1$ circle is given by
\begin{eqnarray}
\nonumber
{\cal W}(\mu_e)&=&e^{i\mu_e\oint\bm A \cdot d\bm \ell}=e^{i\mu_e\int_{y_1}^{y_2} dy\int_{0}^L dz B_1}\\
&&\rightarrow e^{i\mu_e\omega(x,y)}\,,
\end{eqnarray}
where $\mu_e$ is the electric charge of the Wilson line probe and we used (\ref{electric and magnetic fields}). Also, the 't Hooft loop operator that measures the electric flux penetrating the  $y-z$ plane is given by
\begin{eqnarray}
\nonumber
{\cal T}(\mu_m,\theta)&=&e^{-i\mu_m\frac{2\pi}{g^2}\int_{y-z}ds n^1 E_1}\\
&&\rightarrow e^{i\mu_m\left(\sigma(x,y)+\frac{\theta}{2\pi}\omega(x,y)\right)}\,, 
\end{eqnarray}
and $\mu_m$ is the probe magnetic charge.

Now, starting with a pure 't Hooft operator at $\theta=0$, we find upon sending $\theta\rightarrow \theta+2\pi$
\begin{eqnarray}
{\cal T}\left(\mu_m,\theta\rightarrow \theta+2\pi\right)={\cal T}(\mu_m,\theta=0){\cal W}(\mu_m)\,,
\end{eqnarray}
i.e. the magnetic probe acquires an electric charge $\mu_m$. This is Witten's effect  at work.

The main lesson to be learnt here   is that the static monopole considered as the  {\it external source} becomes the electrically charged dyon, in full agreement with the Witten's effect \cite{Witten:1979ey}. However, when the monopoles become the dynamical degrees of freedom and themselves determine the ground state of the ensemble  they remain pure magnetic monopoles as demonstrated in Section \ref{dyons-dynamics}. 

At this point one may wonder how and why the $\theta$ parameter remains to be an observable parameter of the theory when 
exclusively abelian gauge fields are present in the system. Indeed, normally we assume that  the $\theta$ parameter in Maxwell abelian QED is not physical
because the $\theta$ term in Maxwell QED can be expressed as the total derivative which can be removed from the action  due to  the triviality of the  topology. 
The key point relevant for our present discussions is that Witten's effect for the abelian magnetic monopole is operational because the monopole itself determines the nontrivial topology and the $\theta$ parameter becomes the physical parameter in QED in the nontrivial topological (not vacuum) sector determined by the monopole's charge. 

A similar effect when $\theta$ becomes a physically observable parameter also holds for Maxwell QED when the external 
magnetic flux selects a nontrivial topological sector of the system, as argued in  \cite{Cao:2017ocv}. This effect, in fact, represents a novel idea 
on the axion search experiments when the system is sensitive to $\theta$ itself, rather than to $\partial_{\mu}\theta$ as in conventional axion search experiments. 

In the context of the present work these arguments make it clear that the {\it external}  magnetic monopoles become the electrically charge dyons in the presence of $\theta\neq 0$ in the given topological winding sector determined by the external magnetic charge itself. The {\it dynamical} magnetic monopoles remain pure monopoles as they cannot select the topological winding sector for the entire system. Precisely these  {\it dynamical} monopoles condense and determine the ground state of the system. This interpretation is perfectly consistent with our conclusion at the end of section \ref{dyons-dynamics} that the confinement in supersymmetric and non-supersymmetric theories at $\theta\neq 0$ is due to the condensation of the same magnetic monopoles, rather than dyons.

\section*{Conclusion}
The main claim of the  present work can be formulated as follows. We showed that the confinement in the gauge systems with $\theta\neq 0$ is 
a result of  the condensation of the same monopole's configurations which generate the confinement at $\theta=0$.  It should be contrasted  with a conventional lore that
 the confinement at $\theta\neq 0$ is a result of the condensation of the dyons. 

The  $\theta$ parameter is obviously a  physical parameter of the system since all other observables, including  the vacuum energy, are explicitly dependent on $\theta$. Furthermore,  $\cal{CP}$ invariance is explicitly broken 
  for $\theta\neq 0$ as the  computations of the vacuum expectation value  of the topological density (\ref{top}) suggest. However, the  $\theta$ dependence emerges in the system not  as a result of any modifications of any gauge configurations, in comparison with $\theta=0$ case.
  Rather, the $\theta$ dependence emerges in the system    as a result of selection of the specific superposition of the $|\theta, m\ra$ states  as discussed in section \ref{classification}. 
  
  A simple way to interpret this result  is to view the classification $|\theta, m\ra$ in gauge theories in terms of the reduced Brillouin zone scheme as it is normally done in condensed matter physics,  when $\theta$ parameter plays the role of the quasi-momentum in the   $m-$th Brillouin zone. In our classification the parameter $m$ 
  corresponds to the $m-$th metastable state. Using this analogy it is quiet obvious that all the microscopical elements for any $|\theta, m\ra$ states are the same. It is just a specific  selection of the  Bloch type superposition (constructed from  the condensates of $N$ different   of monopole's species)  which provides a complete description of the 
 $|\theta, m\ra$ state. 
 
   We conclude this work with the following short comments.
   It has been recent renewal interests in $\cal{CP}$ invariance of the gauge theories at $\theta=\pi$ 
\cite{Gaiotto:2017yup,Gaiotto:2017tne}. While the questions addressed in \cite{Gaiotto:2017yup,Gaiotto:2017tne} and in our work 
are somewhat different, nevertheless we   observe a number of generic features discussed in  \cite{Gaiotto:2017yup,Gaiotto:2017tne} which have their counterparts in our simplified ``deformed  QCD" model. For example we obviously observe that there is a degeneracy at $\theta=\pi$ in our framework as one can see from classification scheme presented  in section \ref{classification}. Furthermore, one can explicitly see from   (\ref{magnetization}), (\ref{top})  that $\cal{CP}$ invariance is spontaneously broken at $\theta=\pi$, and the sign of $\cal{CP}$ violation  is different depending on the direction this point is approached: $\theta=\pi\pm\epsilon$.  These drastic changes correspond to complete reconstruction of the ground state when  
the system jumps to another Brillouin zone in the  reduced classification scheme 
as described in section \ref{classification}.   
Such a behavior  obviously signals a phase transition at $\theta=\pi$. The superpositions of these two degenerate states at  $\theta=\pi$  can make  $\cal{CP}$ odd and $\cal{CP}$ even ground states. 

One can trace the presence of the metastable states (which eventually become degenerate states at $\theta=\pi$) to the presence of nonlocal operator, the holonomy,  in the system.  Exactly  this feature of non-locality leads to a number of properties  in ``deformed QCD" model which are normally attributed to topologically ordered systems as argued in \cite{Zhitnitsky:2013hs}. Precisely  this sensitivity to arbitrary large distances in gapped theories might be the key element in understanding of the vacuum energy in cosmology because this type of the vacuum energy is generated by non-local physics and cannot be renormalized by any  UV counter-terms, as recently advocated in  \cite{Barvinsky:2017lfl}.

 There are many arguments, presented in section \ref{smooth}, suggesting  that this picture holds in strongly coupled regime as well. Therefore, we strongly believe that in QCD
 we have precisely the same picture for the confinement at $\theta\neq 0$ including metastable states. 
 If this is indeed the case, it may have  
   profound  observational  effects on the  axion production rate   as mentioned in section \ref{phenomenology} due to the nontrivial topological features of the system.  It may be also important for understanding of the nature of the vacuum energy in cosmology   as mentioned above.  It may also affect  the axion domain  wall formation  due to the $2\pi$ periodicity in $\theta$ and presence of the metastable states, see footnote \ref{DW} for references.  The very same metastable states, in general, violate $\cal{CP}$ invariance of the system as they effectively correspond to non-vanishing  $\theta_{\rm eff}=2\pi m/N$. One could speculate \cite{Bhoonah:2014gpa} that precisely these  metastable states might be  responsible for the $\cal{CP}$-odd  correlations observed at RHIC and the LHC.

\section*{Acknowledgements}
We are thankful  to 
Ken Konishi for discussions of the  Witten's effect and its role in oblique confinement in SUSY theories. 
 
The work of M.A. is supported by the NSF grant PHY-1720135 and by Murdock Charitable Trust. The work of A.Z. was supported in part by the Natural Sciences and Engineering Research Council of Canada.

\bibliographystyle{apsrev4-1}

\begin{thebibliography}{63}%
\makeatletter
\providecommand \@ifxundefined [1]{%
 \@ifx{#1\undefined}
}%
\providecommand \@ifnum [1]{%
 \ifnum #1\expandafter \@firstoftwo
 \else \expandafter \@secondoftwo
 \fi
}%
\providecommand \@ifx [1]{%
 \ifx #1\expandafter \@firstoftwo
 \else \expandafter \@secondoftwo
 \fi
}%
\providecommand \natexlab [1]{#1}%
\providecommand \enquote  [1]{``#1''}%
\providecommand \bibnamefont  [1]{#1}%
\providecommand \bibfnamefont [1]{#1}%
\providecommand \citenamefont [1]{#1}%
\providecommand \href@noop [0]{\@secondoftwo}%
\providecommand \href [0]{\begingroup \@sanitize@url \@href}%
\providecommand \@href[1]{\@@startlink{#1}\@@href}%
\providecommand \@@href[1]{\endgroup#1\@@endlink}%
\providecommand \@sanitize@url [0]{\catcode `\\12\catcode `\$12\catcode
  `\&12\catcode `\#12\catcode `\^12\catcode `\_12\catcode `\%12\relax}%
\providecommand \@@startlink[1]{}%
\providecommand \@@endlink[0]{}%
\providecommand \url  [0]{\begingroup\@sanitize@url \@url }%
\providecommand \@url [1]{\endgroup\@href {#1}{\urlprefix }}%
\providecommand \urlprefix  [0]{URL }%
\providecommand \Eprint [0]{\href }%
\providecommand \doibase [0]{http://dx.doi.org/}%
\providecommand \selectlanguage [0]{\@gobble}%
\providecommand \bibinfo  [0]{\@secondoftwo}%
\providecommand \bibfield  [0]{\@secondoftwo}%
\providecommand \translation [1]{[#1]}%
\providecommand \BibitemOpen [0]{}%
\providecommand \bibitemStop [0]{}%
\providecommand \bibitemNoStop [0]{.\EOS\space}%
\providecommand \EOS [0]{\spacefactor3000\relax}%
\providecommand \BibitemShut  [1]{\csname bibitem#1\endcsname}%
\let\auto@bib@innerbib\@empty
\bibitem [{\citenamefont {Witten}(1980)}]{Witten:1980sp}%
  \BibitemOpen
  \bibfield  {author} {\bibinfo {author} {\bibfnamefont {E.}~\bibnamefont
  {Witten}},\ }\href {\doibase 10.1016/0003-4916(80)90325-5} {\bibfield
  {journal} {\bibinfo  {journal} {Annals Phys.}\ }\textbf {\bibinfo {volume}
  {128}},\ \bibinfo {pages} {363} (\bibinfo {year} {1980})}\BibitemShut
  {NoStop}%
\bibitem [{\citenamefont {Witten}(1998)}]{Witten:1998uka}%
  \BibitemOpen
  \bibfield  {author} {\bibinfo {author} {\bibfnamefont {E.}~\bibnamefont
  {Witten}},\ }\href {\doibase 10.1103/PhysRevLett.81.2862} {\bibfield
  {journal} {\bibinfo  {journal} {Phys. Rev. Lett.}\ }\textbf {\bibinfo
  {volume} {81}},\ \bibinfo {pages} {2862} (\bibinfo {year} {1998})},\ \Eprint
  {http://arxiv.org/abs/hep-th/9807109} {arXiv:hep-th/9807109 [hep-th]}
  \BibitemShut {NoStop}%
\bibitem [{\citenamefont {Peccei}\ and\ \citenamefont
  {Quinn}(1977)}]{Peccei:1977ur}%
  \BibitemOpen
  \bibfield  {author} {\bibinfo {author} {\bibfnamefont {R.~D.}\ \bibnamefont
  {Peccei}}\ and\ \bibinfo {author} {\bibfnamefont {H.~R.}\ \bibnamefont
  {Quinn}},\ }\href {\doibase 10.1103/PhysRevD.16.1791} {\bibfield  {journal}
  {\bibinfo  {journal} {Phys. Rev.}\ }\textbf {\bibinfo {volume} {D16}},\
  \bibinfo {pages} {1791} (\bibinfo {year} {1977})}\BibitemShut {NoStop}%
\bibitem [{\citenamefont {Weinberg}(1978)}]{Weinberg:1977ma}%
  \BibitemOpen
  \bibfield  {author} {\bibinfo {author} {\bibfnamefont {S.}~\bibnamefont
  {Weinberg}},\ }\href {\doibase 10.1103/PhysRevLett.40.223} {\bibfield
  {journal} {\bibinfo  {journal} {Phys. Rev. Lett.}\ }\textbf {\bibinfo
  {volume} {40}},\ \bibinfo {pages} {223} (\bibinfo {year} {1978})}\BibitemShut
  {NoStop}%
\bibitem [{\citenamefont {Wilczek}(1978)}]{Wilczek:1977pj}%
  \BibitemOpen
  \bibfield  {author} {\bibinfo {author} {\bibfnamefont {F.}~\bibnamefont
  {Wilczek}},\ }\href {\doibase 10.1103/PhysRevLett.40.279} {\bibfield
  {journal} {\bibinfo  {journal} {Phys. Rev. Lett.}\ }\textbf {\bibinfo
  {volume} {40}},\ \bibinfo {pages} {279} (\bibinfo {year} {1978})}\BibitemShut
  {NoStop}%
\bibitem [{\citenamefont {Shifman}\ \emph {et~al.}(1980)\citenamefont
  {Shifman}, \citenamefont {Vainshtein},\ and\ \citenamefont
  {Zakharov}}]{Shifman:1979if}%
  \BibitemOpen
  \bibfield  {author} {\bibinfo {author} {\bibfnamefont {M.~A.}\ \bibnamefont
  {Shifman}}, \bibinfo {author} {\bibfnamefont {A.~I.}\ \bibnamefont
  {Vainshtein}}, \ and\ \bibinfo {author} {\bibfnamefont {V.~I.}\ \bibnamefont
  {Zakharov}},\ }\href {\doibase 10.1016/0550-3213(80)90209-6} {\bibfield
  {journal} {\bibinfo  {journal} {Nucl. Phys.}\ }\textbf {\bibinfo {volume}
  {B166}},\ \bibinfo {pages} {493} (\bibinfo {year} {1980})}\BibitemShut
  {NoStop}%
\bibitem [{\citenamefont {Kim}(1979)}]{Kim:1979if}%
  \BibitemOpen
  \bibfield  {author} {\bibinfo {author} {\bibfnamefont {J.~E.}\ \bibnamefont
  {Kim}},\ }\href {\doibase 10.1103/PhysRevLett.43.103} {\bibfield  {journal}
  {\bibinfo  {journal} {Phys. Rev. Lett.}\ }\textbf {\bibinfo {volume} {43}},\
  \bibinfo {pages} {103} (\bibinfo {year} {1979})}\BibitemShut {NoStop}%
\bibitem [{\citenamefont {Dine}\ \emph {et~al.}(1981)\citenamefont {Dine},
  \citenamefont {Fischler},\ and\ \citenamefont {Srednicki}}]{Dine:1981rt}%
  \BibitemOpen
  \bibfield  {author} {\bibinfo {author} {\bibfnamefont {M.}~\bibnamefont
  {Dine}}, \bibinfo {author} {\bibfnamefont {W.}~\bibnamefont {Fischler}}, \
  and\ \bibinfo {author} {\bibfnamefont {M.}~\bibnamefont {Srednicki}},\ }\href
  {\doibase 10.1016/0370-2693(81)90590-6} {\bibfield  {journal} {\bibinfo
  {journal} {Phys. Lett.}\ }\textbf {\bibinfo {volume} {104B}},\ \bibinfo
  {pages} {199} (\bibinfo {year} {1981})}\BibitemShut {NoStop}%
\bibitem [{\citenamefont {Zhitnitsky}(1980)}]{Zhitnitsky:1980tq}%
  \BibitemOpen
  \bibfield  {author} {\bibinfo {author} {\bibfnamefont {A.~R.}\ \bibnamefont
  {Zhitnitsky}},\ }\href@noop {} {\bibfield  {journal} {\bibinfo  {journal}
  {Sov. J. Nucl. Phys.}\ }\textbf {\bibinfo {volume} {31}},\ \bibinfo {pages}
  {260} (\bibinfo {year} {1980})},\ \bibinfo {note} {[Yad.
  Fiz.31,497(1980)]}\BibitemShut {NoStop}%
\bibitem [{\citenamefont {van Bibber}\ and\ \citenamefont
  {Rosenberg}(2006)}]{vanBibber:2006rb}%
  \BibitemOpen
  \bibfield  {author} {\bibinfo {author} {\bibfnamefont {K.}~\bibnamefont {van
  Bibber}}\ and\ \bibinfo {author} {\bibfnamefont {L.~J.}\ \bibnamefont
  {Rosenberg}},\ }\href {\doibase 10.1063/1.2349730} {\bibfield  {journal}
  {\bibinfo  {journal} {Phys. Today}\ }\textbf {\bibinfo {volume} {59N8}},\
  \bibinfo {pages} {30} (\bibinfo {year} {2006})}\BibitemShut {NoStop}%
\bibitem [{\citenamefont {Asztalos}\ \emph {et~al.}(2006)\citenamefont
  {Asztalos}, \citenamefont {Rosenberg}, \citenamefont {van Bibber},
  \citenamefont {Sikivie},\ and\ \citenamefont {Zioutas}}]{Asztalos:2006kz}%
  \BibitemOpen
  \bibfield  {author} {\bibinfo {author} {\bibfnamefont {S.~J.}\ \bibnamefont
  {Asztalos}}, \bibinfo {author} {\bibfnamefont {L.~J.}\ \bibnamefont
  {Rosenberg}}, \bibinfo {author} {\bibfnamefont {K.}~\bibnamefont {van
  Bibber}}, \bibinfo {author} {\bibfnamefont {P.}~\bibnamefont {Sikivie}}, \
  and\ \bibinfo {author} {\bibfnamefont {K.}~\bibnamefont {Zioutas}},\ }\href
  {\doibase 10.1146/annurev.nucl.56.080805.140513} {\bibfield  {journal}
  {\bibinfo  {journal} {Ann. Rev. Nucl. Part. Sci.}\ }\textbf {\bibinfo
  {volume} {56}},\ \bibinfo {pages} {293} (\bibinfo {year} {2006})}\BibitemShut
  {NoStop}%
\bibitem [{\citenamefont {Raffelt}(2008)}]{Raffelt:2006cw}%
  \BibitemOpen
  \bibfield  {author} {\bibinfo {author} {\bibfnamefont {G.~G.}\ \bibnamefont
  {Raffelt}},\ }\bibfield  {booktitle} {\emph {\bibinfo {booktitle} {{Axions:
  Theory, cosmology, and experimental searches. Proceedings, 1st Joint
  ILIAS-CERN-CAST axion training, Geneva, Switzerland, November 30-December 2,
  2005}}},\ }\href {\doibase 10.1007/978-3-540-73518-2_3} {\bibfield  {journal}
  {\bibinfo  {journal} {Lect. Notes Phys.}\ }\textbf {\bibinfo {volume}
  {741}},\ \bibinfo {pages} {51} (\bibinfo {year} {2008})},\ \bibinfo {note}
  {[,51(2006)]},\ \Eprint {http://arxiv.org/abs/hep-ph/0611350}
  {arXiv:hep-ph/0611350 [hep-ph]} \BibitemShut {NoStop}%
\bibitem [{\citenamefont {Sikivie}(2010)}]{Sikivie:2009fv}%
  \BibitemOpen
  \bibfield  {author} {\bibinfo {author} {\bibfnamefont {P.}~\bibnamefont
  {Sikivie}},\ }\bibfield  {booktitle} {\emph {\bibinfo {booktitle} {{Crossing
  the boundaries: Gauge dynamics at strong coupling. Proceedings, Workshop in
  Honor of the 60th Birthday of Misha Shifman, Minneapolis, USA, May 14-17,
  2009}}},\ }\href {\doibase 10.1142/S0217751X10048846} {\bibfield  {journal}
  {\bibinfo  {journal} {Int. J. Mod. Phys.}\ }\textbf {\bibinfo {volume}
  {A25}},\ \bibinfo {pages} {554} (\bibinfo {year} {2010})},\ \Eprint
  {http://arxiv.org/abs/0909.0949} {arXiv:0909.0949 [hep-ph]} \BibitemShut
  {NoStop}%
\bibitem [{\citenamefont {Rosenberg}(2015)}]{Rosenberg:2015kxa}%
  \BibitemOpen
  \bibfield  {author} {\bibinfo {author} {\bibfnamefont {L.~J.}\ \bibnamefont
  {Rosenberg}},\ }\bibfield  {booktitle} {\emph {\bibinfo {booktitle} {{Sackler
  Colloquium: Dark Matter Universe: On the Threshold of Discovery Irvine, USA,
  October 18-20, 2012}}},\ }\href {\doibase 10.1073/pnas.1308788112} {\bibfield
   {journal} {\bibinfo  {journal} {Proc. Nat. Acad. Sci.}\ }\textbf {\bibinfo
  {volume} {112}},\ \bibinfo {pages} {12278} (\bibinfo {year}
  {2015})}\BibitemShut {NoStop}%
\bibitem [{\citenamefont {Graham}\ \emph {et~al.}(2015)\citenamefont {Graham},
  \citenamefont {Irastorza}, \citenamefont {Lamoreaux}, \citenamefont
  {Lindner},\ and\ \citenamefont {van Bibber}}]{Graham:2015ouw}%
  \BibitemOpen
  \bibfield  {author} {\bibinfo {author} {\bibfnamefont {P.~W.}\ \bibnamefont
  {Graham}}, \bibinfo {author} {\bibfnamefont {I.~G.}\ \bibnamefont
  {Irastorza}}, \bibinfo {author} {\bibfnamefont {S.~K.}\ \bibnamefont
  {Lamoreaux}}, \bibinfo {author} {\bibfnamefont {A.}~\bibnamefont {Lindner}},
  \ and\ \bibinfo {author} {\bibfnamefont {K.~A.}\ \bibnamefont {van Bibber}},\
  }\href {\doibase 10.1146/annurev-nucl-102014-022120} {\bibfield  {journal}
  {\bibinfo  {journal} {Ann. Rev. Nucl. Part. Sci.}\ }\textbf {\bibinfo
  {volume} {65}},\ \bibinfo {pages} {485} (\bibinfo {year} {2015})},\ \Eprint
  {http://arxiv.org/abs/1602.00039} {arXiv:1602.00039 [hep-ex]} \BibitemShut
  {NoStop}%
\bibitem [{\citenamefont {Ringwald}(2016)}]{Ringwald:2016yge}%
  \BibitemOpen
  \bibfield  {author} {\bibinfo {author} {\bibfnamefont {A.}~\bibnamefont
  {Ringwald}},\ }\bibfield  {booktitle} {\emph {\bibinfo {booktitle}
  {{Proceedings, Neutrino Oscillation Workshop (NOW 2016): Otranto (Lecce),
  Italy, September 4-11, 2016}}},\ }\href@noop {} {\bibfield  {journal}
  {\bibinfo  {journal} {PoS}\ }\textbf {\bibinfo {volume} {NOW2016}},\ \bibinfo
  {pages} {081} (\bibinfo {year} {2016})},\ \Eprint
  {http://arxiv.org/abs/1612.08933} {arXiv:1612.08933 [hep-ph]} \BibitemShut
  {NoStop}%
\bibitem [{\citenamefont {Kitano}\ and\ \citenamefont
  {Yamada}(2015)}]{Kitano:2015fla}%
  \BibitemOpen
  \bibfield  {author} {\bibinfo {author} {\bibfnamefont {R.}~\bibnamefont
  {Kitano}}\ and\ \bibinfo {author} {\bibfnamefont {N.}~\bibnamefont
  {Yamada}},\ }\href {\doibase doi:10.1007/JHEP10(2015)136} {\bibfield
  {journal} {\bibinfo  {journal} {JHEP}\ }\textbf {\bibinfo {volume} {1510}},\
  \bibinfo {pages} {136} (\bibinfo {year} {2015})},\ \Eprint
  {http://arxiv.org/abs/1506.00370} {arXiv:1506.00370 [hep-ph]} \BibitemShut
  {NoStop}%
\bibitem [{\citenamefont {Bonati}\ \emph {et~al.}(2016)\citenamefont {Bonati},
  \citenamefont {D'Elia}, \citenamefont {Mariti}, \citenamefont {Martinelli},
  \citenamefont {Mesiti}, \citenamefont {Negro}, \citenamefont {Sanfilippo},\
  and\ \citenamefont {Villadoro}}]{Bonati:2015vqz}%
  \BibitemOpen
  \bibfield  {author} {\bibinfo {author} {\bibfnamefont {C.}~\bibnamefont
  {Bonati}}, \bibinfo {author} {\bibfnamefont {M.}~\bibnamefont {D'Elia}},
  \bibinfo {author} {\bibfnamefont {M.}~\bibnamefont {Mariti}}, \bibinfo
  {author} {\bibfnamefont {G.}~\bibnamefont {Martinelli}}, \bibinfo {author}
  {\bibfnamefont {M.}~\bibnamefont {Mesiti}}, \bibinfo {author} {\bibfnamefont
  {F.}~\bibnamefont {Negro}}, \bibinfo {author} {\bibfnamefont
  {F.}~\bibnamefont {Sanfilippo}}, \ and\ \bibinfo {author} {\bibfnamefont
  {G.}~\bibnamefont {Villadoro}},\ }\href {\doibase 10.1007/JHEP03(2016)155}
  {\bibfield  {journal} {\bibinfo  {journal} {JHEP}\ }\textbf {\bibinfo
  {volume} {03}},\ \bibinfo {pages} {155} (\bibinfo {year} {2016})},\ \Eprint
  {http://arxiv.org/abs/1512.06746} {arXiv:1512.06746 [hep-lat]} \BibitemShut
  {NoStop}%
\bibitem [{\citenamefont {Borsanyi}\ \emph {et~al.}(2016)\citenamefont
  {Borsanyi} \emph {et~al.}}]{Borsanyi:2016ksw}%
  \BibitemOpen
  \bibfield  {author} {\bibinfo {author} {\bibfnamefont {S.}~\bibnamefont
  {Borsanyi}} \emph {et~al.},\ }\href {\doibase 10.1038/nature20115} {\bibfield
   {journal} {\bibinfo  {journal} {Nature}\ }\textbf {\bibinfo {volume}
  {539}},\ \bibinfo {pages} {69} (\bibinfo {year} {2016})},\ \Eprint
  {http://arxiv.org/abs/1606.07494} {arXiv:1606.07494 [hep-lat]} \BibitemShut
  {NoStop}%
\bibitem [{\citenamefont {Petreczky}\ \emph {et~al.}(2016)\citenamefont
  {Petreczky}, \citenamefont {Schadler},\ and\ \citenamefont
  {Sharma}}]{Petreczky:2016vrs}%
  \BibitemOpen
  \bibfield  {author} {\bibinfo {author} {\bibfnamefont {P.}~\bibnamefont
  {Petreczky}}, \bibinfo {author} {\bibfnamefont {H.-P.}\ \bibnamefont
  {Schadler}}, \ and\ \bibinfo {author} {\bibfnamefont {S.}~\bibnamefont
  {Sharma}},\ }\href {\doibase 10.1016/j.physletb.2016.09.063} {\bibfield
  {journal} {\bibinfo  {journal} {Phys. Lett.}\ }\textbf {\bibinfo {volume}
  {B762}},\ \bibinfo {pages} {498} (\bibinfo {year} {2016})},\ \Eprint
  {http://arxiv.org/abs/1606.03145} {arXiv:1606.03145 [hep-lat]} \BibitemShut
  {NoStop}%
\bibitem [{\citenamefont {D'Elia}\ and\ \citenamefont
  {Negro}(2012)}]{DElia:2012pvq}%
  \BibitemOpen
  \bibfield  {author} {\bibinfo {author} {\bibfnamefont {M.}~\bibnamefont
  {D'Elia}}\ and\ \bibinfo {author} {\bibfnamefont {F.}~\bibnamefont {Negro}},\
  }\href {\doibase 10.1103/PhysRevLett.109.072001} {\bibfield  {journal}
  {\bibinfo  {journal} {Phys. Rev. Lett.}\ }\textbf {\bibinfo {volume} {109}},\
  \bibinfo {pages} {072001} (\bibinfo {year} {2012})},\ \Eprint
  {http://arxiv.org/abs/1205.0538} {arXiv:1205.0538 [hep-lat]} \BibitemShut
  {NoStop}%
\bibitem [{\citenamefont {D'Elia}\ and\ \citenamefont
  {Negro}(2013)}]{DElia:2013uaf}%
  \BibitemOpen
  \bibfield  {author} {\bibinfo {author} {\bibfnamefont {M.}~\bibnamefont
  {D'Elia}}\ and\ \bibinfo {author} {\bibfnamefont {F.}~\bibnamefont {Negro}},\
  }\href {\doibase 10.1103/PhysRevD.88.034503} {\bibfield  {journal} {\bibinfo
  {journal} {Phys. Rev.}\ }\textbf {\bibinfo {volume} {D88}},\ \bibinfo {pages}
  {034503} (\bibinfo {year} {2013})},\ \Eprint {http://arxiv.org/abs/1306.2919}
  {arXiv:1306.2919 [hep-lat]} \BibitemShut {NoStop}%
\bibitem [{\citenamefont {Liang}\ and\ \citenamefont
  {Zhitnitsky}(2016)}]{Liang:2016tqc}%
  \BibitemOpen
  \bibfield  {author} {\bibinfo {author} {\bibfnamefont {X.}~\bibnamefont
  {Liang}}\ and\ \bibinfo {author} {\bibfnamefont {A.}~\bibnamefont
  {Zhitnitsky}},\ }\href {\doibase 10.1103/PhysRevD.94.083502} {\bibfield
  {journal} {\bibinfo  {journal} {Phys. Rev.}\ }\textbf {\bibinfo {volume}
  {D94}},\ \bibinfo {pages} {083502} (\bibinfo {year} {2016})},\ \Eprint
  {http://arxiv.org/abs/1606.00435} {arXiv:1606.00435 [hep-ph]} \BibitemShut
  {NoStop}%
\bibitem [{\citenamefont {Ge}\ \emph {et~al.}(2017)\citenamefont {Ge},
  \citenamefont {Liang},\ and\ \citenamefont {Zhitnitsky}}]{Ge:2017ttc}%
  \BibitemOpen
  \bibfield  {author} {\bibinfo {author} {\bibfnamefont {S.}~\bibnamefont
  {Ge}}, \bibinfo {author} {\bibfnamefont {X.}~\bibnamefont {Liang}}, \ and\
  \bibinfo {author} {\bibfnamefont {A.}~\bibnamefont {Zhitnitsky}},\ }\href
  {\doibase 10.1103/PhysRevD.96.063514} {\bibfield  {journal} {\bibinfo
  {journal} {Phys. Rev.}\ }\textbf {\bibinfo {volume} {D96}},\ \bibinfo {pages}
  {063514} (\bibinfo {year} {2017})},\ \Eprint
  {http://arxiv.org/abs/1702.04354} {arXiv:1702.04354 [hep-ph]} \BibitemShut
  {NoStop}%
\bibitem [{\citenamefont {'t~Hooft}(1981)}]{tHooft:1981bkw}%
  \BibitemOpen
  \bibfield  {author} {\bibinfo {author} {\bibfnamefont {G.}~\bibnamefont
  {'t~Hooft}},\ }\href {\doibase 10.1016/0550-3213(81)90442-9} {\bibfield
  {journal} {\bibinfo  {journal} {Nucl. Phys.}\ }\textbf {\bibinfo {volume}
  {B190}},\ \bibinfo {pages} {455} (\bibinfo {year} {1981})}\BibitemShut
  {NoStop}%
\bibitem [{\citenamefont {Unsal}\ and\ \citenamefont
  {Yaffe}(2008)}]{Unsal:2008ch}%
  \BibitemOpen
  \bibfield  {author} {\bibinfo {author} {\bibfnamefont {M.}~\bibnamefont
  {Unsal}}\ and\ \bibinfo {author} {\bibfnamefont {L.~G.}\ \bibnamefont
  {Yaffe}},\ }\href {\doibase 10.1103/PhysRevD.78.065035} {\bibfield  {journal}
  {\bibinfo  {journal} {Phys. Rev.}\ }\textbf {\bibinfo {volume} {D78}},\
  \bibinfo {pages} {065035} (\bibinfo {year} {2008})},\ \Eprint
  {http://arxiv.org/abs/0803.0344} {arXiv:0803.0344 [hep-th]} \BibitemShut
  {NoStop}%
\bibitem [{\citenamefont {Poppitz}\ and\ \citenamefont
  {Unsal}(2011)}]{Poppitz:2011wy}%
  \BibitemOpen
  \bibfield  {author} {\bibinfo {author} {\bibfnamefont {E.}~\bibnamefont
  {Poppitz}}\ and\ \bibinfo {author} {\bibfnamefont {M.}~\bibnamefont
  {Unsal}},\ }\href {\doibase 10.1007/JHEP07(2011)082} {\bibfield  {journal}
  {\bibinfo  {journal} {JHEP}\ }\textbf {\bibinfo {volume} {07}},\ \bibinfo
  {pages} {082} (\bibinfo {year} {2011})},\ \Eprint
  {http://arxiv.org/abs/1105.3969} {arXiv:1105.3969 [hep-th]} \BibitemShut
  {NoStop}%
\bibitem [{\citenamefont {Thomas}\ and\ \citenamefont
  {Zhitnitsky}(2012)}]{Thomas:2011ee}%
  \BibitemOpen
  \bibfield  {author} {\bibinfo {author} {\bibfnamefont {E.}~\bibnamefont
  {Thomas}}\ and\ \bibinfo {author} {\bibfnamefont {A.~R.}\ \bibnamefont
  {Zhitnitsky}},\ }\href {\doibase 10.1103/PhysRevD.85.044039} {\bibfield
  {journal} {\bibinfo  {journal} {Phys. Rev.}\ }\textbf {\bibinfo {volume}
  {D85}},\ \bibinfo {pages} {044039} (\bibinfo {year} {2012})},\ \Eprint
  {http://arxiv.org/abs/1109.2608} {arXiv:1109.2608 [hep-th]} \BibitemShut
  {NoStop}%
\bibitem [{\citenamefont {Anber}\ \emph {et~al.}(2012)\citenamefont {Anber},
  \citenamefont {Poppitz},\ and\ \citenamefont {Unsal}}]{Anber:2011gn}%
  \BibitemOpen
  \bibfield  {author} {\bibinfo {author} {\bibfnamefont {M.~M.}\ \bibnamefont
  {Anber}}, \bibinfo {author} {\bibfnamefont {E.}~\bibnamefont {Poppitz}}, \
  and\ \bibinfo {author} {\bibfnamefont {M.}~\bibnamefont {Unsal}},\ }\href
  {\doibase 10.1007/JHEP04(2012)040} {\bibfield  {journal} {\bibinfo  {journal}
  {JHEP}\ }\textbf {\bibinfo {volume} {04}},\ \bibinfo {pages} {040} (\bibinfo
  {year} {2012})},\ \Eprint {http://arxiv.org/abs/1112.6389} {arXiv:1112.6389
  [hep-th]} \BibitemShut {NoStop}%
\bibitem [{\citenamefont {Unsal}(2012)}]{Unsal:2012zj}%
  \BibitemOpen
  \bibfield  {author} {\bibinfo {author} {\bibfnamefont {M.}~\bibnamefont
  {Unsal}},\ }\href {\doibase 10.1103/PhysRevD.86.105012} {\bibfield  {journal}
  {\bibinfo  {journal} {Phys. Rev.}\ }\textbf {\bibinfo {volume} {D86}},\
  \bibinfo {pages} {105012} (\bibinfo {year} {2012})},\ \Eprint
  {http://arxiv.org/abs/1201.6426} {arXiv:1201.6426 [hep-th]} \BibitemShut
  {NoStop}%
\bibitem [{\citenamefont {Poppitz}\ \emph {et~al.}(2012)\citenamefont
  {Poppitz}, \citenamefont {Sch{\"a}fer},\ and\ \citenamefont
  {Unsal}}]{Poppitz:2012sw}%
  \BibitemOpen
  \bibfield  {author} {\bibinfo {author} {\bibfnamefont {E.}~\bibnamefont
  {Poppitz}}, \bibinfo {author} {\bibfnamefont {T.}~\bibnamefont
  {Sch{\"a}fer}}, \ and\ \bibinfo {author} {\bibfnamefont {M.}~\bibnamefont
  {Unsal}},\ }\href {\doibase 10.1007/JHEP10(2012)115} {\bibfield  {journal}
  {\bibinfo  {journal} {JHEP}\ }\textbf {\bibinfo {volume} {10}},\ \bibinfo
  {pages} {115} (\bibinfo {year} {2012})},\ \Eprint
  {http://arxiv.org/abs/1205.0290} {arXiv:1205.0290 [hep-th]} \BibitemShut
  {NoStop}%
\bibitem [{\citenamefont {Thomas}\ and\ \citenamefont
  {Zhitnitsky}(2013)}]{Thomas:2012tu}%
  \BibitemOpen
  \bibfield  {author} {\bibinfo {author} {\bibfnamefont {E.}~\bibnamefont
  {Thomas}}\ and\ \bibinfo {author} {\bibfnamefont {A.~R.}\ \bibnamefont
  {Zhitnitsky}},\ }\href {\doibase 10.1103/PhysRevD.87.085027} {\bibfield
  {journal} {\bibinfo  {journal} {Phys. Rev.}\ }\textbf {\bibinfo {volume}
  {D87}},\ \bibinfo {pages} {085027} (\bibinfo {year} {2013})},\ \Eprint
  {http://arxiv.org/abs/1208.2030} {arXiv:1208.2030 [hep-ph]} \BibitemShut
  {NoStop}%
\bibitem [{\citenamefont {Poppitz}\ \emph {et~al.}(2013)\citenamefont
  {Poppitz}, \citenamefont {Sch{\"a}fer},\ and\ \citenamefont
  {{\"U}nsal}}]{Poppitz:2012nz}%
  \BibitemOpen
  \bibfield  {author} {\bibinfo {author} {\bibfnamefont {E.}~\bibnamefont
  {Poppitz}}, \bibinfo {author} {\bibfnamefont {T.}~\bibnamefont
  {Sch{\"a}fer}}, \ and\ \bibinfo {author} {\bibfnamefont {M.}~\bibnamefont
  {{\"U}nsal}},\ }\href {\doibase 10.1007/JHEP03(2013)087} {\bibfield
  {journal} {\bibinfo  {journal} {JHEP}\ }\textbf {\bibinfo {volume} {03}},\
  \bibinfo {pages} {087} (\bibinfo {year} {2013})},\ \Eprint
  {http://arxiv.org/abs/1212.1238} {arXiv:1212.1238 [hep-th]} \BibitemShut
  {NoStop}%
\bibitem [{\citenamefont {Zhitnitsky}(2013)}]{Zhitnitsky:2013hs}%
  \BibitemOpen
  \bibfield  {author} {\bibinfo {author} {\bibfnamefont {A.~R.}\ \bibnamefont
  {Zhitnitsky}},\ }\href {\doibase 10.1016/j.aop.2013.05.020} {\bibfield
  {journal} {\bibinfo  {journal} {Annals Phys.}\ }\textbf {\bibinfo {volume}
  {336}},\ \bibinfo {pages} {462} (\bibinfo {year} {2013})},\ \Eprint
  {http://arxiv.org/abs/1301.7072} {arXiv:1301.7072 [hep-ph]} \BibitemShut
  {NoStop}%
\bibitem [{\citenamefont {Anber}(2013)}]{Anber:2013sga}%
  \BibitemOpen
  \bibfield  {author} {\bibinfo {author} {\bibfnamefont {M.~M.}\ \bibnamefont
  {Anber}},\ }\href {\doibase 10.1103/PhysRevD.88.085003} {\bibfield  {journal}
  {\bibinfo  {journal} {Phys. Rev.}\ }\textbf {\bibinfo {volume} {D88}},\
  \bibinfo {pages} {085003} (\bibinfo {year} {2013})},\ \Eprint
  {http://arxiv.org/abs/1302.2641} {arXiv:1302.2641 [hep-th]} \BibitemShut
  {NoStop}%
\bibitem [{\citenamefont {Anber}\ \emph {et~al.}(2013)\citenamefont {Anber},
  \citenamefont {Collier}, \citenamefont {Poppitz}, \citenamefont
  {Strimas-Mackey},\ and\ \citenamefont {Teeple}}]{Anber:2013doa}%
  \BibitemOpen
  \bibfield  {author} {\bibinfo {author} {\bibfnamefont {M.~M.}\ \bibnamefont
  {Anber}}, \bibinfo {author} {\bibfnamefont {S.}~\bibnamefont {Collier}},
  \bibinfo {author} {\bibfnamefont {E.}~\bibnamefont {Poppitz}}, \bibinfo
  {author} {\bibfnamefont {S.}~\bibnamefont {Strimas-Mackey}}, \ and\ \bibinfo
  {author} {\bibfnamefont {B.}~\bibnamefont {Teeple}},\ }\href {\doibase
  10.1007/JHEP11(2013)142} {\bibfield  {journal} {\bibinfo  {journal} {JHEP}\
  }\textbf {\bibinfo {volume} {11}},\ \bibinfo {pages} {142} (\bibinfo {year}
  {2013})},\ \Eprint {http://arxiv.org/abs/1310.3522} {arXiv:1310.3522
  [hep-th]} \BibitemShut {NoStop}%
\bibitem [{\citenamefont {Anber}\ \emph {et~al.}(2014)\citenamefont {Anber},
  \citenamefont {Poppitz},\ and\ \citenamefont {Teeple}}]{Anber:2014lba}%
  \BibitemOpen
  \bibfield  {author} {\bibinfo {author} {\bibfnamefont {M.~M.}\ \bibnamefont
  {Anber}}, \bibinfo {author} {\bibfnamefont {E.}~\bibnamefont {Poppitz}}, \
  and\ \bibinfo {author} {\bibfnamefont {B.}~\bibnamefont {Teeple}},\ }\href
  {\doibase 10.1007/JHEP09(2014)040} {\bibfield  {journal} {\bibinfo  {journal}
  {JHEP}\ }\textbf {\bibinfo {volume} {09}},\ \bibinfo {pages} {040} (\bibinfo
  {year} {2014})},\ \Eprint {http://arxiv.org/abs/1406.1199} {arXiv:1406.1199
  [hep-th]} \BibitemShut {NoStop}%
\bibitem [{\citenamefont {Anber}\ and\ \citenamefont
  {Vincent-Genod}(2017)}]{Anber:2017pak}%
  \BibitemOpen
  \bibfield  {author} {\bibinfo {author} {\bibfnamefont {M.~M.}\ \bibnamefont
  {Anber}}\ and\ \bibinfo {author} {\bibfnamefont {L.}~\bibnamefont
  {Vincent-Genod}},\ }\href@noop {} {\  (\bibinfo {year} {2017})},\ \Eprint
  {http://arxiv.org/abs/1704.08277} {arXiv:1704.08277 [hep-th]} \BibitemShut
  {NoStop}%
\bibitem [{\citenamefont {Bhoonah}\ \emph {et~al.}(2014)\citenamefont
  {Bhoonah}, \citenamefont {Thomas},\ and\ \citenamefont
  {Zhitnitsky}}]{Bhoonah:2014gpa}%
  \BibitemOpen
  \bibfield  {author} {\bibinfo {author} {\bibfnamefont {A.}~\bibnamefont
  {Bhoonah}}, \bibinfo {author} {\bibfnamefont {E.}~\bibnamefont {Thomas}}, \
  and\ \bibinfo {author} {\bibfnamefont {A.~R.}\ \bibnamefont {Zhitnitsky}},\
  }\href {\doibase 10.1016/j.nuclphysb.2014.11.007} {\bibfield  {journal}
  {\bibinfo  {journal} {Nucl. Phys.}\ }\textbf {\bibinfo {volume} {B890}},\
  \bibinfo {pages} {30} (\bibinfo {year} {2014})},\ \Eprint
  {http://arxiv.org/abs/1407.5121} {arXiv:1407.5121 [hep-ph]} \BibitemShut
  {NoStop}%
\bibitem [{\citenamefont {Witten}(1979{\natexlab{a}})}]{Witten:1979vv}%
  \BibitemOpen
  \bibfield  {author} {\bibinfo {author} {\bibfnamefont {E.}~\bibnamefont
  {Witten}},\ }\href {\doibase 10.1016/0550-3213(79)90031-2} {\bibfield
  {journal} {\bibinfo  {journal} {Nucl. Phys.}\ }\textbf {\bibinfo {volume}
  {B156}},\ \bibinfo {pages} {269} (\bibinfo {year}
  {1979}{\natexlab{a}})}\BibitemShut {NoStop}%
\bibitem [{\citenamefont {Veneziano}(1979)}]{Veneziano:1979ec}%
  \BibitemOpen
  \bibfield  {author} {\bibinfo {author} {\bibfnamefont {G.}~\bibnamefont
  {Veneziano}},\ }\href {\doibase 10.1016/0550-3213(79)90332-8} {\bibfield
  {journal} {\bibinfo  {journal} {Nucl. Phys.}\ }\textbf {\bibinfo {volume}
  {B159}},\ \bibinfo {pages} {213} (\bibinfo {year} {1979})}\BibitemShut
  {NoStop}%
\bibitem [{\citenamefont {Di~Vecchia}\ and\ \citenamefont
  {Veneziano}(1980)}]{DiVecchia:1980yfw}%
  \BibitemOpen
  \bibfield  {author} {\bibinfo {author} {\bibfnamefont {P.}~\bibnamefont
  {Di~Vecchia}}\ and\ \bibinfo {author} {\bibfnamefont {G.}~\bibnamefont
  {Veneziano}},\ }\href {\doibase 10.1016/0550-3213(80)90370-3} {\bibfield
  {journal} {\bibinfo  {journal} {Nucl. Phys.}\ }\textbf {\bibinfo {volume}
  {B171}},\ \bibinfo {pages} {253} (\bibinfo {year} {1980})}\BibitemShut
  {NoStop}%
\bibitem [{\citenamefont {Witten}(1979{\natexlab{b}})}]{Witten:1979ey}%
  \BibitemOpen
  \bibfield  {author} {\bibinfo {author} {\bibfnamefont {E.}~\bibnamefont
  {Witten}},\ }\href {\doibase 10.1016/0370-2693(79)90838-4} {\bibfield
  {journal} {\bibinfo  {journal} {Phys. Lett.}\ }\textbf {\bibinfo {volume}
  {86B}},\ \bibinfo {pages} {283} (\bibinfo {year}
  {1979}{\natexlab{b}})}\BibitemShut {NoStop}%
\bibitem [{\citenamefont {Shifman}(2012)}]{Shifman:2012zz}%
  \BibitemOpen
  \bibfield  {author} {\bibinfo {author} {\bibfnamefont {M.}~\bibnamefont
  {Shifman}},\ }\href
  {http://www.cambridge.org/mw/academic/subjects/physics/theoretical-physics-and-mathematical-physics/advanced-topics-quantum-field-theory-lecture-course?format=AR}
  {\emph {\bibinfo {title} {{Advanced topics in quantum field theory.}}}}\
  (\bibinfo  {publisher} {Cambridge Univ. Press},\ \bibinfo {address}
  {Cambridge, UK},\ \bibinfo {year} {2012})\BibitemShut {NoStop}%
\bibitem [{\citenamefont {Seiberg}\ and\ \citenamefont
  {Witten}(1994)}]{Seiberg:1994rs}%
  \BibitemOpen
  \bibfield  {author} {\bibinfo {author} {\bibfnamefont {N.}~\bibnamefont
  {Seiberg}}\ and\ \bibinfo {author} {\bibfnamefont {E.}~\bibnamefont
  {Witten}},\ }\href {\doibase 10.1016/0550-3213(94)90124-4,
  10.1016/0550-3213(94)00449-8} {\bibfield  {journal} {\bibinfo  {journal}
  {Nucl. Phys.}\ }\textbf {\bibinfo {volume} {B426}},\ \bibinfo {pages} {19}
  (\bibinfo {year} {1994})},\ \bibinfo {note} {[Erratum: Nucl.
  Phys.B430,485(1994)]},\ \Eprint {http://arxiv.org/abs/hep-th/9407087}
  {arXiv:hep-th/9407087 [hep-th]} \BibitemShut {NoStop}%
\bibitem [{\citenamefont {Weinberg}\ and\ \citenamefont
  {Yi}(2007)}]{Weinberg:2006rq}%
  \BibitemOpen
  \bibfield  {author} {\bibinfo {author} {\bibfnamefont {E.~J.}\ \bibnamefont
  {Weinberg}}\ and\ \bibinfo {author} {\bibfnamefont {P.}~\bibnamefont {Yi}},\
  }\href {\doibase 10.1016/j.physrep.2006.11.002} {\bibfield  {journal}
  {\bibinfo  {journal} {Phys. Rept.}\ }\textbf {\bibinfo {volume} {438}},\
  \bibinfo {pages} {65} (\bibinfo {year} {2007})},\ \Eprint
  {http://arxiv.org/abs/hep-th/0609055} {arXiv:hep-th/0609055 [hep-th]}
  \BibitemShut {NoStop}%
\bibitem [{\citenamefont {Julia}\ and\ \citenamefont
  {Zee}(1975)}]{Julia:1975ff}%
  \BibitemOpen
  \bibfield  {author} {\bibinfo {author} {\bibfnamefont {B.}~\bibnamefont
  {Julia}}\ and\ \bibinfo {author} {\bibfnamefont {A.}~\bibnamefont {Zee}},\
  }\href {\doibase 10.1103/PhysRevD.11.2227} {\bibfield  {journal} {\bibinfo
  {journal} {Phys. Rev.}\ }\textbf {\bibinfo {volume} {D11}},\ \bibinfo {pages}
  {2227} (\bibinfo {year} {1975})}\BibitemShut {NoStop}%
\bibitem [{\citenamefont {Bak}\ \emph {et~al.}(1998)\citenamefont {Bak},
  \citenamefont {Lee},\ and\ \citenamefont {Lee}}]{Bak:1997pc}%
  \BibitemOpen
  \bibfield  {author} {\bibinfo {author} {\bibfnamefont {D.}~\bibnamefont
  {Bak}}, \bibinfo {author} {\bibfnamefont {C.-k.}\ \bibnamefont {Lee}}, \ and\
  \bibinfo {author} {\bibfnamefont {K.-M.}\ \bibnamefont {Lee}},\ }\href
  {\doibase 10.1103/PhysRevD.57.5239} {\bibfield  {journal} {\bibinfo
  {journal} {Phys. Rev.}\ }\textbf {\bibinfo {volume} {D57}},\ \bibinfo {pages}
  {5239} (\bibinfo {year} {1998})},\ \Eprint
  {http://arxiv.org/abs/hep-th/9708149} {arXiv:hep-th/9708149 [hep-th]}
  \BibitemShut {NoStop}%
\bibitem [{\citenamefont {Di~Pierro}\ and\ \citenamefont
  {Konishi}(1996)}]{DiPierro:1996zj}%
  \BibitemOpen
  \bibfield  {author} {\bibinfo {author} {\bibfnamefont {M.}~\bibnamefont
  {Di~Pierro}}\ and\ \bibinfo {author} {\bibfnamefont {K.}~\bibnamefont
  {Konishi}},\ }\href {\doibase 10.1016/0370-2693(96)01136-7} {\bibfield
  {journal} {\bibinfo  {journal} {Phys. Lett.}\ }\textbf {\bibinfo {volume}
  {B388}},\ \bibinfo {pages} {90} (\bibinfo {year} {1996})},\ \Eprint
  {http://arxiv.org/abs/hep-th/9605178} {arXiv:hep-th/9605178 [hep-th]}
  \BibitemShut {NoStop}%
\bibitem [{\citenamefont {Konishi}(1997)}]{Konishi:1996iz}%
  \BibitemOpen
  \bibfield  {author} {\bibinfo {author} {\bibfnamefont {K.}~\bibnamefont
  {Konishi}},\ }\bibfield  {booktitle} {\emph {\bibinfo {booktitle} {{High
  energy physics: Proceedings, 28th International Conference, ICHEP'96, Warsaw,
  Poland, July 25-31, 1996. Vol. 1, 2}}},\ }\href {\doibase
  10.1016/S0370-2693(96)01527-4} {\bibfield  {journal} {\bibinfo  {journal}
  {Phys. Lett.}\ }\textbf {\bibinfo {volume} {B392}},\ \bibinfo {pages} {101}
  (\bibinfo {year} {1997})},\ \Eprint {http://arxiv.org/abs/hep-th/9609021}
  {arXiv:hep-th/9609021 [hep-th]} \BibitemShut {NoStop}%
\bibitem [{\citenamefont {Konishi}\ and\ \citenamefont
  {Terao}(1998)}]{Konishi:1998mk}%
  \BibitemOpen
  \bibfield  {author} {\bibinfo {author} {\bibfnamefont {K.}~\bibnamefont
  {Konishi}}\ and\ \bibinfo {author} {\bibfnamefont {H.}~\bibnamefont
  {Terao}},\ }\href {\doibase 10.1016/S0550-3213(97)00739-6} {\bibfield
  {journal} {\bibinfo  {journal} {Nucl. Phys.}\ }\textbf {\bibinfo {volume}
  {B511}},\ \bibinfo {pages} {264} (\bibinfo {year} {1998})},\ \Eprint
  {http://arxiv.org/abs/hep-th/9707005} {arXiv:hep-th/9707005 [hep-th]}
  \BibitemShut {NoStop}%
\bibitem [{\citenamefont {Chen}\ \emph {et~al.}(2010)\citenamefont {Chen},
  \citenamefont {Dorey},\ and\ \citenamefont {Petunin}}]{Chen:2010yr}%
  \BibitemOpen
  \bibfield  {author} {\bibinfo {author} {\bibfnamefont {H.-Y.}\ \bibnamefont
  {Chen}}, \bibinfo {author} {\bibfnamefont {N.}~\bibnamefont {Dorey}}, \ and\
  \bibinfo {author} {\bibfnamefont {K.}~\bibnamefont {Petunin}},\ }\href
  {\doibase 10.1007/JHEP06(2010)024} {\bibfield  {journal} {\bibinfo  {journal}
  {JHEP}\ }\textbf {\bibinfo {volume} {06}},\ \bibinfo {pages} {024} (\bibinfo
  {year} {2010})},\ \Eprint {http://arxiv.org/abs/1004.0703} {arXiv:1004.0703
  [hep-th]} \BibitemShut {NoStop}%
\bibitem [{\citenamefont {Chen}\ \emph {et~al.}(2011)\citenamefont {Chen},
  \citenamefont {Dorey},\ and\ \citenamefont {Petunin}}]{Chen:2011gk}%
  \BibitemOpen
  \bibfield  {author} {\bibinfo {author} {\bibfnamefont {H.-Y.}\ \bibnamefont
  {Chen}}, \bibinfo {author} {\bibfnamefont {N.}~\bibnamefont {Dorey}}, \ and\
  \bibinfo {author} {\bibfnamefont {K.}~\bibnamefont {Petunin}},\ }\href
  {\doibase 10.1007/JHEP11(2011)020} {\bibfield  {journal} {\bibinfo  {journal}
  {JHEP}\ }\textbf {\bibinfo {volume} {11}},\ \bibinfo {pages} {020} (\bibinfo
  {year} {2011})},\ \Eprint {http://arxiv.org/abs/1105.4584} {arXiv:1105.4584
  [hep-th]} \BibitemShut {NoStop}%
\bibitem [{\citenamefont {Dorey}(2001)}]{Dorey:2000dt}%
  \BibitemOpen
  \bibfield  {author} {\bibinfo {author} {\bibfnamefont {N.}~\bibnamefont
  {Dorey}},\ }\href {\doibase 10.1088/1126-6708/2001/04/008} {\bibfield
  {journal} {\bibinfo  {journal} {JHEP}\ }\textbf {\bibinfo {volume} {04}},\
  \bibinfo {pages} {008} (\bibinfo {year} {2001})},\ \Eprint
  {http://arxiv.org/abs/hep-th/0010115} {arXiv:hep-th/0010115 [hep-th]}
  \BibitemShut {NoStop}%
\bibitem [{\citenamefont {Dorey}\ and\ \citenamefont
  {Parnachev}(2001)}]{Dorey:2000qc}%
  \BibitemOpen
  \bibfield  {author} {\bibinfo {author} {\bibfnamefont {N.}~\bibnamefont
  {Dorey}}\ and\ \bibinfo {author} {\bibfnamefont {A.}~\bibnamefont
  {Parnachev}},\ }\href {\doibase 10.1088/1126-6708/2001/08/059} {\bibfield
  {journal} {\bibinfo  {journal} {JHEP}\ }\textbf {\bibinfo {volume} {08}},\
  \bibinfo {pages} {059} (\bibinfo {year} {2001})},\ \Eprint
  {http://arxiv.org/abs/hep-th/0011202} {arXiv:hep-th/0011202 [hep-th]}
  \BibitemShut {NoStop}%
\bibitem [{\citenamefont {Liu}\ \emph {et~al.}(2015{\natexlab{a}})\citenamefont
  {Liu}, \citenamefont {Shuryak},\ and\ \citenamefont {Zahed}}]{Liu:2015ufa}%
  \BibitemOpen
  \bibfield  {author} {\bibinfo {author} {\bibfnamefont {Y.}~\bibnamefont
  {Liu}}, \bibinfo {author} {\bibfnamefont {E.}~\bibnamefont {Shuryak}}, \ and\
  \bibinfo {author} {\bibfnamefont {I.}~\bibnamefont {Zahed}},\ }\href
  {\doibase 10.1103/PhysRevD.92.085006} {\bibfield  {journal} {\bibinfo
  {journal} {Phys. Rev.}\ }\textbf {\bibinfo {volume} {D92}},\ \bibinfo {pages}
  {085006} (\bibinfo {year} {2015}{\natexlab{a}})},\ \Eprint
  {http://arxiv.org/abs/1503.03058} {arXiv:1503.03058 [hep-ph]} \BibitemShut
  {NoStop}%
\bibitem [{\citenamefont {Liu}\ \emph {et~al.}(2015{\natexlab{b}})\citenamefont
  {Liu}, \citenamefont {Shuryak},\ and\ \citenamefont {Zahed}}]{Liu:2015jsa}%
  \BibitemOpen
  \bibfield  {author} {\bibinfo {author} {\bibfnamefont {Y.}~\bibnamefont
  {Liu}}, \bibinfo {author} {\bibfnamefont {E.}~\bibnamefont {Shuryak}}, \ and\
  \bibinfo {author} {\bibfnamefont {I.}~\bibnamefont {Zahed}},\ }\href
  {\doibase 10.1103/PhysRevD.92.085007} {\bibfield  {journal} {\bibinfo
  {journal} {Phys. Rev.}\ }\textbf {\bibinfo {volume} {D92}},\ \bibinfo {pages}
  {085007} (\bibinfo {year} {2015}{\natexlab{b}})},\ \Eprint
  {http://arxiv.org/abs/1503.09148} {arXiv:1503.09148 [hep-ph]} \BibitemShut
  {NoStop}%
\bibitem [{\citenamefont {Larsen}\ and\ \citenamefont
  {Shuryak}(2015)}]{Larsen:2015vaa}%
  \BibitemOpen
  \bibfield  {author} {\bibinfo {author} {\bibfnamefont {R.}~\bibnamefont
  {Larsen}}\ and\ \bibinfo {author} {\bibfnamefont {E.}~\bibnamefont
  {Shuryak}},\ }\href {\doibase 10.1103/PhysRevD.92.094022} {\bibfield
  {journal} {\bibinfo  {journal} {Phys. Rev.}\ }\textbf {\bibinfo {volume}
  {D92}},\ \bibinfo {pages} {094022} (\bibinfo {year} {2015})},\ \Eprint
  {http://arxiv.org/abs/1504.03341} {arXiv:1504.03341 [hep-ph]} \BibitemShut
  {NoStop}%
\bibitem [{\citenamefont {Anber}\ and\ \citenamefont
  {Poppitz}(2015)}]{Anber:2015wha}%
  \BibitemOpen
  \bibfield  {author} {\bibinfo {author} {\bibfnamefont {M.~M.}\ \bibnamefont
  {Anber}}\ and\ \bibinfo {author} {\bibfnamefont {E.}~\bibnamefont
  {Poppitz}},\ }\href {\doibase 10.1007/JHEP10(2015)051} {\bibfield  {journal}
  {\bibinfo  {journal} {JHEP}\ }\textbf {\bibinfo {volume} {10}},\ \bibinfo
  {pages} {051} (\bibinfo {year} {2015})},\ \Eprint
  {http://arxiv.org/abs/1508.00910} {arXiv:1508.00910 [hep-th]} \BibitemShut
  {NoStop}%
\bibitem [{\citenamefont {Cao}\ and\ \citenamefont
  {Zhitnitsky}(2017)}]{Cao:2017ocv}%
  \BibitemOpen
  \bibfield  {author} {\bibinfo {author} {\bibfnamefont {C.}~\bibnamefont
  {Cao}}\ and\ \bibinfo {author} {\bibfnamefont {A.}~\bibnamefont
  {Zhitnitsky}},\ }\href {\doibase 10.1103/PhysRevD.96.015013} {\bibfield
  {journal} {\bibinfo  {journal} {Phys. Rev.}\ }\textbf {\bibinfo {volume}
  {D96}},\ \bibinfo {pages} {015013} (\bibinfo {year} {2017})},\ \Eprint
  {http://arxiv.org/abs/1702.00012} {arXiv:1702.00012 [hep-ph]} \BibitemShut
  {NoStop}%
\bibitem [{\citenamefont {Gaiotto}\ \emph
  {et~al.}(2017{\natexlab{a}})\citenamefont {Gaiotto}, \citenamefont
  {Kapustin}, \citenamefont {Komargodski},\ and\ \citenamefont
  {Seiberg}}]{Gaiotto:2017yup}%
  \BibitemOpen
  \bibfield  {author} {\bibinfo {author} {\bibfnamefont {D.}~\bibnamefont
  {Gaiotto}}, \bibinfo {author} {\bibfnamefont {A.}~\bibnamefont {Kapustin}},
  \bibinfo {author} {\bibfnamefont {Z.}~\bibnamefont {Komargodski}}, \ and\
  \bibinfo {author} {\bibfnamefont {N.}~\bibnamefont {Seiberg}},\ }\href
  {\doibase 10.1007/JHEP05(2017)091} {\bibfield  {journal} {\bibinfo  {journal}
  {JHEP}\ }\textbf {\bibinfo {volume} {05}},\ \bibinfo {pages} {091} (\bibinfo
  {year} {2017}{\natexlab{a}})},\ \Eprint {http://arxiv.org/abs/1703.00501}
  {arXiv:1703.00501 [hep-th]} \BibitemShut {NoStop}%
\bibitem [{\citenamefont {Gaiotto}\ \emph
  {et~al.}(2017{\natexlab{b}})\citenamefont {Gaiotto}, \citenamefont
  {Komargodski},\ and\ \citenamefont {Seiberg}}]{Gaiotto:2017tne}%
  \BibitemOpen
  \bibfield  {author} {\bibinfo {author} {\bibfnamefont {D.}~\bibnamefont
  {Gaiotto}}, \bibinfo {author} {\bibfnamefont {Z.}~\bibnamefont
  {Komargodski}}, \ and\ \bibinfo {author} {\bibfnamefont {N.}~\bibnamefont
  {Seiberg}},\ }\href@noop {} {\  (\bibinfo {year} {2017}{\natexlab{b}})},\
  \Eprint {http://arxiv.org/abs/1708.06806} {arXiv:1708.06806 [hep-th]}
  \BibitemShut {NoStop}%
\bibitem [{\citenamefont {Barvinsky}\ and\ \citenamefont
  {Zhitnitsky}(2017)}]{Barvinsky:2017lfl}%
  \BibitemOpen
  \bibfield  {author} {\bibinfo {author} {\bibfnamefont {A.~O.}\ \bibnamefont
  {Barvinsky}}\ and\ \bibinfo {author} {\bibfnamefont {A.~R.}\ \bibnamefont
  {Zhitnitsky}},\ }\href@noop {} {\  (\bibinfo {year} {2017})},\ \Eprint
  {http://arxiv.org/abs/1709.09671} {arXiv:1709.09671 [hep-th]} \BibitemShut
  {NoStop}%
\end{thebibliography}
%

\end{document}